\documentclass[11pt]{article}
\pdfoutput=1
\usepackage{graphicx}
\usepackage{enumerate}
\usepackage{epsfig}
\usepackage{epstopdf}
\usepackage{bm} 

\usepackage{amsfonts,amssymb,amsmath}
\usepackage{latexsym}
\usepackage[english]{babel}

\newcommand{\dd}{{\rm d}}

\newcommand{\be}{\begin{equation}}
\newcommand{\beq}{\begin{equation}}
\newcommand{\eeq}{\end{equation}}
\newcommand{\eq}{\begin{equation}}
\newcommand{\ee}{\end{equation}}
\newcommand{\eqn}{\begin{eqnarray}}
\newcommand{\eeqn}{\end{eqnarray}}
\newcommand{\arr}{\begin{eqnarray*}}
\newcommand{\earr}{\end{eqnarray*}}

\newcommand{\mc}[1]{{\mathcal {#1}}}
\newcommand{\GN}{{\mc G}}

\newcommand{\A}{{\cal A}}

\newcommand{\eg}{{\it e.g.}\ }
\newcommand{\ie}{{\it i.e.}\ } 
\newcommand{\p}{\partial}

\newcommand{\la}{\lambda}
\newcommand{\si}{\sigma}

\newcommand{\lp}{\left(}
\newcommand{\rp}{\right)}
\newcommand{\lb}{\left[}
\newcommand{\rb}{\right]}

\begin{document}
\setlength{\unitlength}{1mm}

\thispagestyle{empty}
\begin{flushright}
\small \tt
\begin{tabular}{l}
CPHT-RR004.0211\\
LPT-ORSAY 11-14
\end{tabular}
\end{flushright}
\vspace*{1.cm}

\begin{center}
{\bf \LARGE  Black funnels and droplets in thermal equilibrium }\\

\vspace*{1.5cm}

{\bf Marco M.~Caldarelli,}$^{1}\,$ {\bf \'Oscar J.C.~Dias,}$^{2}\,$
{\bf Ricardo Monteiro,}$^{3}\,$ {\bf Jorge E.~Santos}$^{4}\,$

\vspace*{1cm}

{\it $^1$\,Laboratoire de Physique Th\'eorique, Univ. Paris-Sud,\\
 CNRS UMR 8627,
 F-91405 Orsay, France\\
 and\\
Centre de Physique Th\'eorique, Ecole Polytechnique,\\
 CNRS UMR 7644,
 F-91128 Palaiseau, France}\\[.3em]

%
{\it $^2\,$ DAMTP, Centre for Mathematical Sciences, University of Cambridge,\\
 Wilberforce Road, Cambridge CB3 0WA, United Kingdom}\\[.3em]
{\it $^3\,$ The Niels Bohr International Academy, The Niels Bohr Institute, \\
Blegdamsvej 17, DK-2100 Copenhagen, Denmark}\\[.3em]
{\it $^4\,$ Department of Physics, UCSB, Santa Barbara, CA 93106, USA}\\[.3em]

\vspace*{0.3cm} {\tt marco.caldarelli@th.u-psud.fr, O.Dias@damtp.cam.ac.uk, \\
monteiro@nbi.dk, jss55@physics.ucsb.edu}

\vspace*{1cm}

\vspace{.5cm} {\bf ABSTRACT}
\vspace{.5cm}

\end{center}

It has recently been proposed that the strong coupling behaviour of
quantum field theories on a non-dynamical black hole background can
be described, in the context of the AdS/CFT correspondence, by a
competition between two gravity duals: a black funnel and a black
droplet. We present here thermal equilibrium solutions which
represent such spacetimes, providing the first example where the
thermal competition between the gravity duals can be studied. The
solutions correspond to a special family of charged AdS C-metrics.
We compute the corresponding Euclidean actions and find that the
black funnel always dominates the canonical ensemble in our example,
meaning that the field theory does not undergo a phase transition.

\noindent


 \setcounter{page}{0} \setcounter{footnote}{0}
\newpage

\tableofcontents


\setcounter{equation}{0}\section{Introduction}

Black holes are simple and highly symmetric solutions of general
relativity,  whose classical behavior is fairly well understood. In
a more general framework, they are located at the interface between
classical and quantum realms, and for this reason they stand out in
the quest for a quantum theory of gravity. The most striking and
well-known effect due to quantum theory is that black holes --
classically perfect absorbers -- evaporate through the emission of
thermal Hawking radiation \cite{Hawking:1974sw}.

Unfortunately, the calculations leading to the Hawking radiation are
quite complicated;  they can be carried out in general only for free
fields propagating on the curved background of the black hole, and
weakly interacting fields in perturbation theory, see for example
\cite{Birrell:1982ix,Ross:2005sc}. Very little is known on the other
hand when the fields interact strongly, and the best one can do is
to define a Hartle-Hawking state using a Euclidean path integral
periodic in imaginary time.

A new road to explore these phenomena has been opened in a series of
articles by  Hubeny, Marolf and Rangamani
\cite{Hubeny:2009ru,Hubeny:2009kz,Hubeny:2009rc} using the power of
the AdS/CFT correspondence
\cite{Maldacena:1997re,Gubser:1998bc,Witten:1998qj}, that maps
strongly coupled quantum field theories to the classical
gravitational dynamics in an higher dimensional AdS spacetime. The
state describing the quantum field theory at strong coupling
propagating on a non-dynamical manifold $\cal M$ is holographically
described by an AdS supergravity solution whose timelike boundary is
conformal to $\cal M$. Ideally, one would like to find the bulk
geometry whose boundary is conformal to the Schwarzschild spacetime.
Such a solution would describe Hawking radiation at strong coupling.
It is however challenging to obtain such a solution without resorting to numerical work, and for this
reason simplified models were initially analyzed: in
\cite{Hubeny:2009ru,Hubeny:2009kz}, Hubey, Marolf and Rangamani studied three-dimensional black holes
appearing on the conformal boundary of the AdS C-metric, while in \cite{Hubeny:2009rc} the problem was
approached by the same authors for AdS black hole backgrounds.

\begin{figure}[htb]
\centerline{\includegraphics{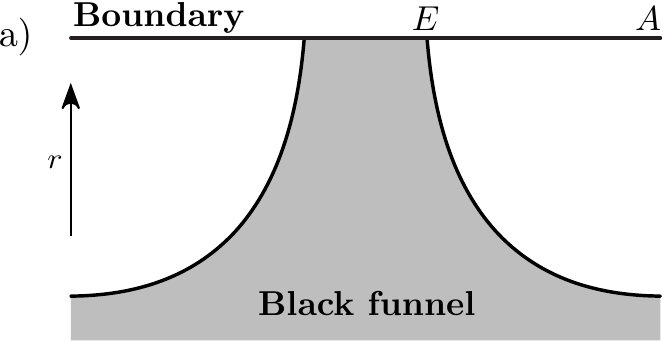}
\hspace{1.5cm}\includegraphics{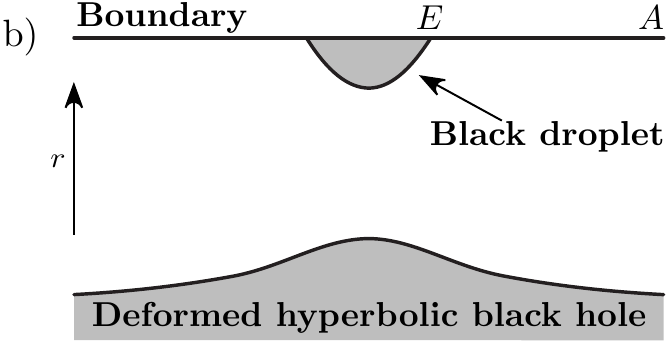}\quad}
\caption{{\bf a)} Schematic spatial representation of a black
funnel. It has a throat that starts at the boundary black hole, and
then extends into the bulk where, at the funnel ``shoulders", it
approaches the geometry of a hyperbolic black hole. Coordinate $r$
is the holographic radial coordinate. The holographic boundary is at
$r\rightarrow \infty$. Point $E$ signals the horizon of the boundary
black hole. Point $A$ describes the asymptotic region of the
holographic boundary. {\bf b)} Black droplet system with two
disconnected horizons. One of them, that joins the horizon of the
boundary black hole, describes a black droplet. The other one, deep
in the bulk, describes a deformed hyperbolic black hole that
approaches the geometry of a hyperbolic black hole for large values
of the radial boundary coordinate.} \label{fig:dropletfunnel}
\end{figure}

Traditionally, the (AdS) C-metric is taken to describe a pair of
uniformly  accelerated black holes, the acceleration being driven by
a string that pulls the black hole (see, e.g.,
\cite{Plebanski:1976gy,Podolsky:2002nk,Dias:2002mi,Krtous:2005ej}).
The authors of \cite{Hubeny:2009ru} showed that a different
interpretation is possible, for which the conformal timelike
boundary is a spacetime containing a black hole, whose event horizon
is continued in the bulk of the solution as shown in
Fig.~\ref{fig:dropletfunnel}. The event horizon of the bulk solution
can be singly connected, and asymptote to the event horizon of a
black brane (Fig.~\ref{fig:dropletfunnel}.a): in that case, the
solution describes a {\em black funnel}. Alternatively, the event
horizon of the boundary black hole can close off in the bulk,
forming a {\em black droplet} floating over a planar black brane
that lives deep in the bulk of the dual spacetime, as shown in
Fig.~\ref{fig:dropletfunnel}.b. These two situations arise in two
different regimes of the Hawking radiation, the former when the
boundary black hole couples strongly with the field theory plasma,
while the latter, due to the lack of connectivity of the two
horizons, describes a situation where the boundary black hole
polarizes the vacuum near the horizon, but couples only weakly to
the field theory plasma dual to the black brane.

So, for each boundary black hole
there are two possible bulk solutions that have the same conformal
timelike boundary, one describing a black funnel, and the other a
black droplet. Both these solutions form gravitational instantons
that contribute to the full Euclidean action of the boundary black
hole and compete to dominate the thermodynamical ensemble. In
\cite{Hubeny:2009ru}, a phase transition was conjectured between
funnel dominated ensembles and black droplet dominated ensembles as
the size $R_{\sf bdy}$ of the {\it asymptotically flat} boundary
black hole and the the distance $R_{\sf S}$ between the ``shoulder"
of the planar black brane or black funnel and the holographic
boundary are tuned. In the neutral system of \cite{Hubeny:2009ru}
the distance $R_{\sf S}$ scales with the inverse of the temperature,
$R_{\sf S}\sim 1/T_H$. More precisely, one should expect the
dynamics of small black holes with $R_{\sf bdy}T_H\ll1$ to be
driven by black droplets, while black funnels should describe the
physics of large, $R_{\sf bdy}T_H\gg1$, asymptotically flat boundary
black holes. This could however not be tested by the authors of
\cite{Hubeny:2009ru}, because the uncharged C-metric can never be in
thermal equilibrium; the black droplet and the planar black hole
have different temperatures and one cannot compare the free energies
of the black funnel and the black droplet.

In this article, we want to explore these ideas further by
considering a more general solution, the charged AdS C-metric
\cite{Plebanski:1976gy,Dias:2002mi} solution of the four dimensional
Einstein-Maxwell theory with negative cosmological constant. Having
at our disposal one extra parameter $-$ the charge of the black hole
$-$ we can tweak it to reach thermal equilibrium. This is seen to
happen in the case of ``lukewarm'' black droplets
\cite{MellorMoss}--\cite{Dias:2004rz}. However, the extra parameter
is not enough to make the chemical potentials of the two event
horizons of the black droplet spacetime to match. Such a match is a
necessary condition for the thermodynamic equilibrium in the
grand-canonical ensemble. Therefore, we shall consider only the
canonical ensemble. Another important point is that, when comparing
the free energies, we will work only with the magnetically charged
solutions. While the charged AdS C-metric allows for electric and/or
magnetic charge, the electromagnetic potential in the electric case
cannot be made everywhere regular in the Euclidean section.

The paper is organised as follows. In
Section~\ref{sec:DropletFunnelEquilibrium}, we will find the
solutions in thermal equilibrium and study their geometry.
Section~\ref{sec:FreeEtensor} will be devoted to the computation of
the Euclidean action for the corresponding black funnel and
``lukewarm" black droplet solutions. This will allow us to compare
the free energies and show that there is no sign of a phase
transition: the black funnels are always the dominant instanton.
This can be understood by the fact that the black droplet and the
planar black hole have charges of opposite sign, which has a cost in
terms of greater free energy of the total system. We conclude in
Section~\ref{sec:AdSCFT}, where we compute the expectation value of
the stress tensor of the quantum fields propagating on the boundary
black hole background, and we discuss the quantum field theory
interpretation of the black funnels and droplets. Finally, in the
appendices we first give a detailed analysis of the charged AdS
C-metric parameter space (Appendix~\ref{sec:NoEquilibrium}),
followed by the analysis of the solutions with horizons in thermal
equilibrium (Appendix~\ref{appB}) and a symmetry of the lukewarm
solutions that simplifies their study.

\setcounter{equation}{0}
\section{Black droplets and funnels in thermal equilibrium
\label{sec:DropletFunnelEquilibrium}}

\renewcommand{\labelenumi}{\Roman{enumi}.}
\renewcommand{\labelenumii}{\Roman{enumi}.\Alph{enumii}.}

 \subsection{The charged AdS
C-metric \label{sec:Cmetric}}
The charged AdS C-metric \cite{Plebanski:1976gy} is a Petrov type D
solution  of four-dimensional Einstein-Maxwell theory in presence of
a cosmological constant $\Lambda=-3/\ell^2$, and has been
extensively studied in recent years
\cite{Podolsky:2002nk,Dias:2002mi,Krtous:2005ej} in many different contexts
\cite{Dias:2003up}--\cite{GPbook}. Its line element can be written as
 \beq ds^2=\frac{1}{\mathcal{A}^2(x-y)^2} \lp -F(y)
\mathcal{A}^2 \ell^2 dt^2
+\frac{dy^2}{F(y)}+\frac{dx^2}{G(x)}+G(x)d\phi^2\rp ,
\label{AdSCmetric}
 \eeq
with fourth order polynomials $F$ and $G$ defined by
 \beq G(x)=1-\kappa\,x^2-2\mu\, x^3-(q_e^2+q_m^2) x^4\,,\qquad
 F(y)=\frac{1}{\mathcal{A}^2\ell^2}-G(y)\,.
\label{AdSCmetricAUX}
 \eeq
The associated Maxwell potential is given by
 \beq A=-q_e \ell \,y\,dt+ \frac{q_m}{\mathcal{A}}\, x\,d\phi\,.
\label{AdSCmetricMaxwell}
 \eeq
and the metric and gauge field satisfy the field equations
 \eqn
 && \nabla_\mu F^{\mu\nu}=0\,,
\nonumber \\
 && R_{\mu\nu}-\frac{1}{2}R\,g_{\mu\nu}+{\Lambda} g_{\mu\nu}=8\pi\GN\,
T_{\mu\nu}^{\mathrm{Max}}\,, \quad \hbox{with}\quad
T_{\mu\nu}^{\mathrm{Max}} = \frac{1}{4\pi\GN}\lp
F_\mu^{\phantom{\alpha}\alpha}F_{\nu\alpha}-\frac{1}{4}g_{\mu\nu}F_{\alpha\beta}F^{\alpha\beta}
\rp \label{eom}
 \eeqn
being the Maxwell energy-momentum tensor associated with the field
strength $F=dA$.

This solution is characterized by six parameters: the cosmological
constant $\Lambda$, the acceleration parameter $\mathcal{A}$, the
mass parameter $\mu$, the gauge charges parameters $q_e$ (electric)
and $q_m$ (magnetic), and the parameter $\kappa$ which determines
the topology of the black holes ($\kappa=1,0,-1$ for horizons with
spherical, planar and hyperbolic topology, respectively).  The
geometry is asymptotically locally anti-de~Sitter (AdS), the
asymptotic region being in the neighborhood of the $x=y$
hypersurface, and has curvature singularities at $y=\pm \infty$ and
at $x=\pm \infty$.

If we choose appropriately the allowed range of the $(x,y)$
coordinates, the AdS C-metric describes a regular spacetime with
horizons clothing the curvature singularities. To guarantee that the
geometry has the correct Lorentzian signature $(-,+,+,+)$, we must
demand that the range of $x$ is such that $G(x)\geq0$, and since
$G(x)<0$ for large $x$, it follows that the coordinate $x$ is
restricted to vary between two roots $x_i$ and $x_j$ of the
polynomial $G(x)$. These roots signal the presence of axes of
symmetry of the solution, corresponding to the rotational symmetry
generated by $\p_\phi$. Typically, the C-metric develops conical
singularities at these poles. To avoid the conical singularity at
one of the poles, say $x_i$, we choose the period of $\phi$ to be
given by
\begin{equation}
\Delta \phi=\frac{4 \pi}{|G'(x_i)|}\:,
\label{PeriodPhi}
\end{equation}
but, this leaves a
conical singularity at the other pole with deficit angle given by
\beq
\delta_j = 2\pi \left ( 1-\frac{G'(x_j)}{\left|G'(x_i)\right|}\right ).
\label{deficit}\eeq

A crucial feature for us is that the asymptotic boundary conditions
that  one sets on the $x=y$ hypersurface can lead to two different
external solutions for the field equations, one obtained in the
region $y\leq x$ and the other for $y \geq x$. The range of $y$ is
therefore chosen to be $y\leq x$ (or $y\geq x$); the upper (lower)
bound $y=x$ describes the asymptotic AdS region and $y=-\infty$ (or
$y=+\infty$) corresponds to a curvature singularity.

The roots of $F(y)$ usually describe the horizons of the
C-metric solution. These can be up to four, labelled $y_i$ with
$i=0\ldots3$, and ordered according to $y_0\leq y_1\leq y_2<y_3$.
The Hawking temperature $T_{H_i}$ associated to the event horizons is
given by their surface gravity evaluated on them, divided by $2\pi$,
\ie
$T_{H_i}^2= -\lp 8\pi^2 \rp^{-1}(\nabla_\mu \xi_\nu)(\nabla^\mu
\xi^\nu){\bigl |}_{H_i}$,
 where $\xi=\partial_t$ is the null generator
of the Killing horizon $H_i$. For the C-metric, this gives
\beq
T_{H_i}=\frac{\mathcal{A\ell}|F^\prime(y_i)|}{4\pi}\,.
\label{beta}\eeq


In the standard interpretation of the C-metric as a pair of
uniformly accelerated black holes, the solution parameters are
usually strongly restricted to guarantee the existence of a continuous
limit as the acceleration parameter goes to zero, where one recovers
the AdS charged black holes. For example, for $\kappa=1$, the range
of $x$ is constrained to be such that after taking the coordinate
transformation $y=-1/(\mathcal{A}r)$ and $x=\cos\theta$ and setting
$\mathcal{A}=0$, one recovers the standard AdS Reissner-Nordstr\"{o}m
black hole with $-1 \leq x \leq 1$. The horizons at $y_0$ and $y_1$
are then, respectively, the inner and outer black hole horizons and
$y=y_3$ is the acceleration horizon. One avoids the conical
singularity at $x=-1$ by choosing the period of $\phi$ to be given
by \eqref{PeriodPhi} and the deficit angle that is left at $x=1$
is interpreted to be sourced by a string, with mass density $\mu_s
=\delta/(8\pi)$ and pressure $p=-\mu_s$, that accelerates the
black hole.

 \subsection{A special family of lukewarm AdS C-metrics  \label{sec:LukeCmetric}}

In the present study, we want to look into the C-metric from a
different perspective. Instead of interpreting it as a pair of
accelerated black holes we want to discuss it as a solution
that describes black droplets and black funnels. The idea is to
extend the analysis of \cite{Hubeny:2009ru,Hubeny:2009kz} to the
charged case, with the advantage that the presence of the charge
allows us to have solutions in thermal equilibrium. This interpretation
demands imposing different requirements from the standard ones, discussed above, on the range of the solution parameters. We will do a
detailed survey over the several possible C-metric solutions in
Section \ref{sec:NoEquilibrium}. However, in this section, we will
focus on the most interesting case for our purposes.

\begin{figure}[t]
\centerline{\includegraphics[width=.35\textwidth]{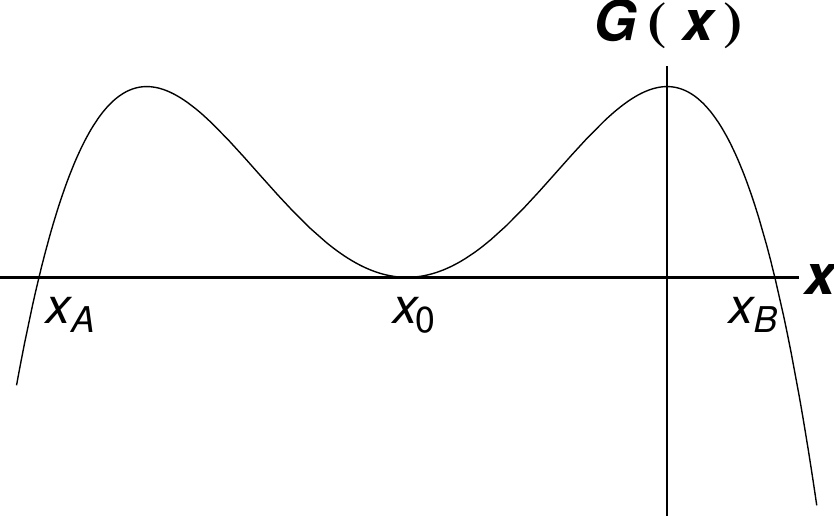}
\hspace{2cm}\includegraphics[width=.35\textwidth]{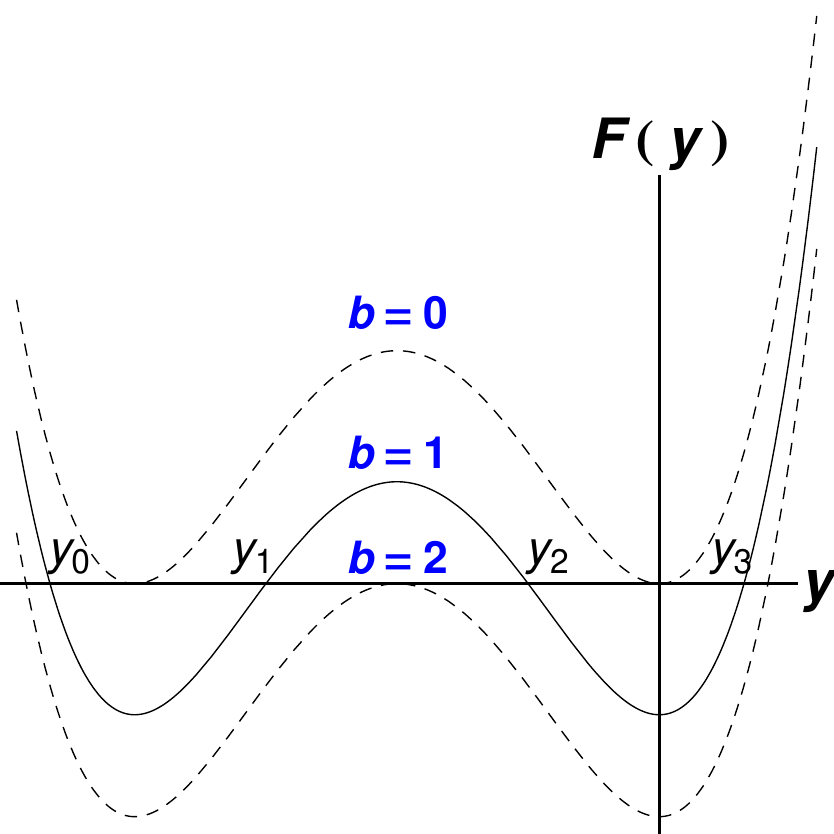}}
\caption{Functions $G(x)$ and $F(y)$ for the lukewarm charged AdS
C-metric with a double zero in $G(x)$: see
\eqref{lukeDoubleProperties}. As explained in the text, the double root $x_0=-2$ of $G(x)$ is at an infinite proper distance from any other point with $x\neq x_0$. There are therefore two disjoint regions of interest where $G(x)>0$: $x_A\leq x\leq x_0$ and $x_0\leq x\leq x_B$, with
$x_A=-2(\sqrt{2}+1)$ and $x_B=2(\sqrt{2}-1)$. On the right panel, we show how the graph of $F(y)$ gets shifted as the parameter $b$ varies in its range $0\leq b\leq 2$. It has four zeros
if $0<b<2$; two zeros if $b=0$; and three zeros if $b=2$. The
horizons at $y_1\equiv-a=b-4$ and $y_2\equiv-b$ have the same
temperature in all these cases.} \label{fig:Lukewarm}
\end{figure}

As motivated in the introduction, we are interested in solutions
that describe a black droplet in thermal equilibrium with a deformed
hyperbolic black hole in the bulk. This translates into finding families
of parameters of the AdS C-metric that have:
\begin{enumerate}[i)]
\item two non-degenerate event horizons at the same temperature;
\item a regular electromagnetic field;
\item a boundary black hole geometry with a well defined asymptotic region.
\end{enumerate}
As will be explained in more detail later, solutions satisfying condition iii) have a double root in the function $G(x)$ defined in \eqref{AdSCmetricAUX}.
Families of solutions
with these properties are described by functions $F(y)$ and $G(x)$
that have the behaviour sketched in Fig. \ref{fig:Lukewarm}. As for condition ii), we mean not only that the Maxwell tensor $F$ is regular, but also that the electromagnetic potential $A$ (in the Euclideanised solution) can be put in a gauge where it is explicitly regular. We will for now neglect
this restriction, and will address it in Section~\ref{sec:FreeEtensor}.

Condition i) requires that the temperatures (\ref{beta}) of the two
event horizons coincide. In the charged AdS C-metric there are four
families of solutions where two of the horizons have the same
temperature. In the literature, these are denoted by the Nariai,
cold, ultracold and lukewarm solutions
\cite{Dias:2004rz}.\footnote{In the standard interpretation of the
C-metric as a pair of accelerated black holes, the Nariai solution
describes the case where the outer black hole horizon $y_1$
coincides with the acceleration horizon $y_2$; in the cold solution
the inner ($y_0$) and outer ($y_1$) black hole horizons coincide;
and the ultracold solution is the limiting case where $y_0=y_1=y_2$.
The cold and ultracold solutions have zero temperature horizons,
justifying their name. In the lukewarm solution, as the name
suggests, the two horizons have the same non-vanishing temperature.
This nomenclature was first introduced by Mellor and Moss
\cite{MellorMoss}, and Romans \cite{Romans} for de Sitter solutions
(see also \cite{Cardoso:2004uz} for a generalization of these
solutions to higher dimensions). The acceleration horizon in the
C-metric solution plays in many respects a similar role as the
cosmological horizon in the de Sitter black holes. Thus it is not a
surprise that solutions with horizons at equal temperature share a
similar structure in the two systems and that the same nomenclature
is used \cite{Mann:1995vb}--\cite{Dias:2004rz}.} The former three
families require zero temperature horizons and we discard them. The
lukewarm solution is the only one that interests us. It describes a
solution where the horizons located at $y=y_1$ and $y=y_2$ have a
non-vanishing temperature. Moreover, in general these horizons do
not coincide, and their temperature is finite. On the other hand,
condition iii) implies that the real zeros of $G$ satisfy $x_A<x_0=
x_1<x_B$ (see Fig. \ref{fig:Lukewarm}.a).

The general analysis of the lukewarm solutions is performed in
Appendix~\ref{appB}, where it is shown that conditions i) and iii)
are satisfied for a one parameter (for fixed $\ell$ and fixed $q_e$
or $q_m$) subfamily of the AdS C-metric, with spherical horizons
($\kappa=+1$) and with the property that the mass parameter $\mu$ and total charge parameter $\sqrt{q_e^2+q_m^2}$ are equal. In short, this solution
satisfies
 \beq
F(y_1)=F(y_2)=0\,, \quad F^\prime(y_1)=-F^\prime(y_2)\,; \qquad
\hbox{and} \quad G(x_0)=0\,, \quad G^\prime(x_0)=0\,,
\label{lukeGconditions}
 \eeq
where $y_1\equiv -a$, $y_2\equiv -b$ and $x_0$ are the roots
displayed in Fig. \ref{fig:Lukewarm}. The conditions on $G(x_0)$ and
$G^\prime(x_0)$ require that $a=4-b$.
 Explicitly, the lukewarm family of solutions that satisfies \eqref{lukeGconditions} is then given
by
 \eqn && F(y)=\frac{1}{16}\left[ y^2(y+4)^2-b^2(b-4)^2\right], \qquad G(x)= 1-\frac{1}{16}\, x^2(x+4)^2\,,
 \nonumber \\
&& \kappa=1\,, \qquad \frac{1}{\mathcal{A}^2\ell^2}=1-\frac{1}{16}\,
b^2(b-4)^2\,, \qquad \mu=\sqrt{q_e^2+q_m^2}=\frac{1}{4}\,, \qquad
0\leq b\leq 2\,.
 \label{lukeDoubleProperties}
 \eeqn
Positivity of $1/\A^2\ell^2$ restricts $b$ to range in
$]-2(\sqrt{2}-1),2(\sqrt{2}+1)[$, the boundary points and $b=2$  corresponding to
the flat spacetime limit. Besides, $b\mapsto4-b$ and
$b\mapsto2-\sqrt{4+4b-b^2}\,$ leave the solution invariant, hence
the range $0\leq b\leq 2$ describes the fundamental domain of the
solution and covers all the cases. We will therefore focus on that interval in the rest of the discussion. For $0<b<2$, $F(y)$ has the four
zeros, $y_0\leq b-4 < -b \leq y_3$. In the limiting case $b=0$,
$F(y)$ has only two degenerate zeros in $y=-4$ and $y=0$. It
corresponds to having the critical case $\mathcal{A}=1/\ell$ where the
cosmological attraction is balanced by the acceleration
$\mathcal{A}$. In the other limiting case, $b=2$, the horizons at
$y=b-4$ and at $y=-b$ coalesce and $F(y)$ has three zeros. It
corresponds to the flat case limit where $\ell\rightarrow\infty$ and
$F(y)=-G(y)$ (see Fig. \ref{fig:Lukewarm}.b).

A property of the $F$ and $G$ polynomials is that, as
shown in Appendix~\ref{appB}, they become even under reflections
through the $x=x_0$ and $y=x_0$ axes, respectively, once the thermal
equilibrium condition is satisfied. Geometrically, this translates
into an isometry that maps the $\{x-y<0;\,x<x_0\}$ region to the
$\{x-y>0;\,x>x_0\}$ region, and the $\{x-y<0;\,x>x_0\}$ region to
the $\{x-y>0;\,x<x_0\}$ region (see appendix~\ref{appB}). Hence, we can
restrict the study of this spacetime to half of the coordinate
space, for example $x>x_0$. For this reason, henceforth, we will
consider the lukewarm solution only in the range $ x_0\leq x\leq
x_B$ and $y_1\leq y\leq y_2$ of the coordinates, corresponding to
two static outer regions with AdS asymptotics as we shall shortly
see (Fig.~\ref{fig:DiagDropFun}).

The horizons at $y=b-4$ and $y=-b$  have the same temperature $T_H=1/\beta$ with \beq
\beta=\frac{4\pi\sqrt{4+4b-b^2}}{(4-b)b}\,. \label{betaFinal}
 \eeq
Moreover, the solution $\eqref{lukeDoubleProperties}$ has no conical
singularities at $x=x_B$ (nor at $x=x_A$) if we take the period of $\phi$ to be
 \beq
\Delta\phi=\sqrt{2}\,\pi \,. \label{PeriodPhiFinal}
 \eeq
Note that with this choice the solution is regular everywhere.
Contrary to the standard claim, we thus have an explicit example of
a charged AdS C-metric that is well behaved everywhere with horizons
clothing the curvature singularity and no conical singularities. In
the neutral case this observation was first pointed out in
\cite{Hubeny:2009ru,Hubeny:2009kz}.

 \subsection{Black droplets and funnels in thermal equilibrium \label{sec:LukeDroplet}}

To show that the solutions studied in the previous subsection describe lukewarm black droplets and black funnels, we need to discuss first some properties of the system.

\begin{figure}
\centerline{\includegraphics[width=.54\textwidth]{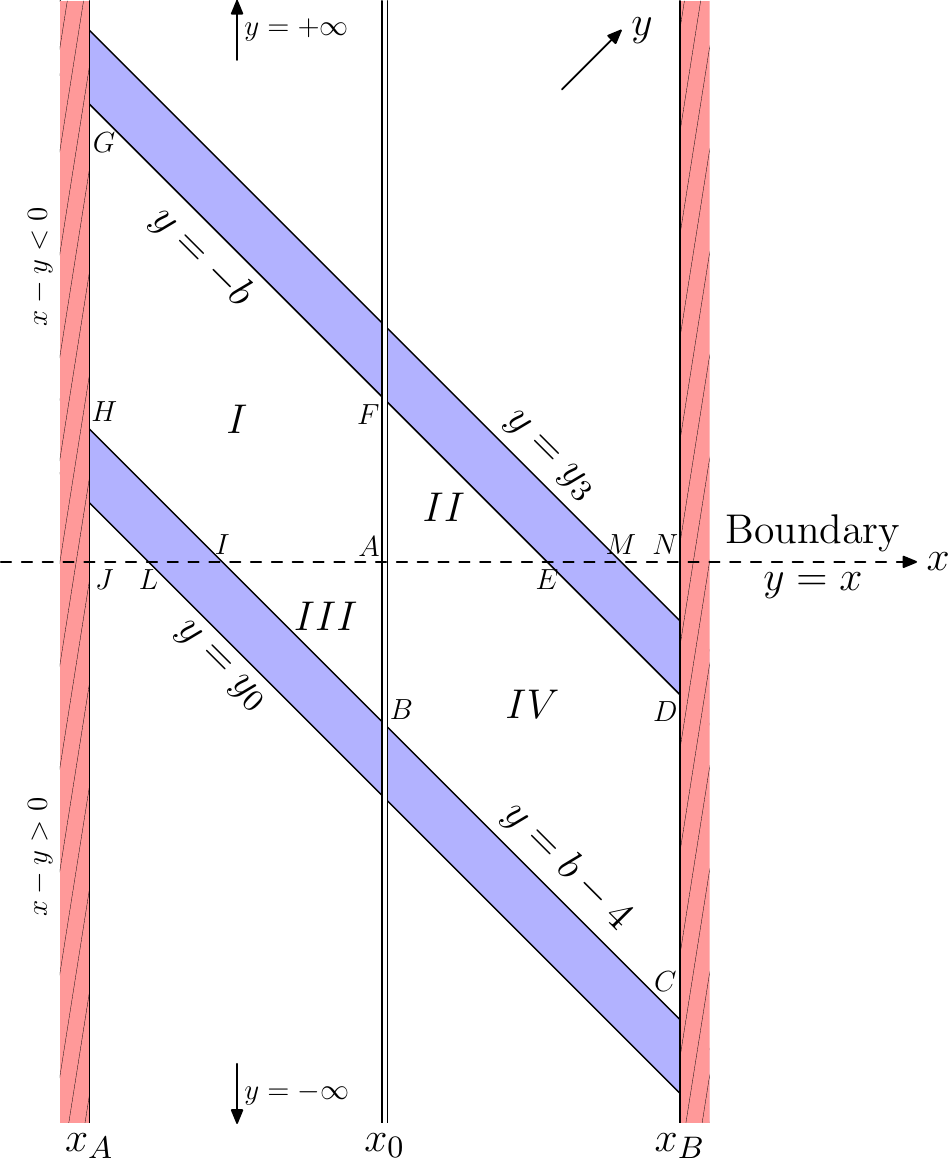}}
\caption{Black droplets (quadrants $I$ and $IV$) and black funnels
(quadrants $II$ and $III$) described by the lukewarm charged AdS
C-metric \eqref{AdSCmetric}, \eqref{AdSCmetricMaxwell} and
\eqref{lukeDoubleProperties}. The black funnel of quadrant $II$ and
the black droplet of quadrant $IV$ share the same boundary at $y=x$
and their outer horizons are at the same temperature. They are
related through a symmetry transformation the solutions of quadrants
$III$ and $I$. The holographic boundary point $A$ is at an infinite
proper distance from any other point with $\chi\neq x_0$ and thus describes the
asymptotic region of the boundary. The white regions are static,
while the colored diagonal strips are between the inner and outer horizons.}
\label{fig:DiagDropFun}
\end{figure}

The range of the coordinates $x$ and $y$ is displayed in
Fig.~\ref{fig:DiagDropFun}. As discussed in the previous paragraph,
quadrants $I$ and $IV$ are isometric, and so are $II$ and $III$. As
we shall see, the points at $x=x_0$ are at infinite proper distance
from any other point in the manifold, and we can therefore restrict
the analysis to quadrants $II$ and $IV$ without loss of generality.
Henceforth we will restrict to the range $x_0\leq x\leq x_B$, the
upper bound due to the fact that we must demand $G(x)\geq0$ to have
solutions with the right Lorentzian signature. The range $x>x_B$ represents the interior of the boundary black hole.

It is useful to rewrite the lukewarm solutions in Fefferman-Graham
coordinates $(r,\chi)$,  both to highlight the structure of their
boundary and to study them in the context of AdS/CFT. This is a
difficult task, but for our purpose  it will be good enough to
define $(r,\chi)$ close to the boundary; for the current solutions
we find that they are obtained by the coordinate transformation
\begin{eqnarray}
y=x+\varepsilon\,\frac{1}{r}\,,\qquad
x=\chi-\varepsilon\,\frac{\mathcal{A}^2\ell^2 G(\chi)}{r} \,, \qquad
\hbox{with}\quad \left\{
\begin{array}{ll}
\varepsilon = -1 \qquad \hbox{if}\!\! & \qquad x-y>0 \,, \\
\varepsilon = +1 \qquad \hbox{if}\!\! & \qquad x-y<0
\,.\label{FGcoords}
\end{array}
\right.
\end{eqnarray}
Here, $r$ is the holographic radial coordinate. Indeed, there is a
curvature singularity at $r=0$ (\ie $y=\pm \infty$) and the
asymptotic AdS boundary is at $r\rightarrow \infty$ (\ie $x=y$). The
coordinate transformation in the polar coordinate, $x(\chi,r)$ was
chosen to guarantee that the cross term between the radial and polar
component of the metric vanishes at infinity up to order $1/r^2$,
\ie $g_{ri}\sim \mathcal{O}(1/r^2)$ as $r\rightarrow\infty$ (with
$i$ being a spacelike boundary coordinate). These correspond to a
perturbative version of the familiar Fefferman-Graham coordinates.
As will be discussed later, this choice guarantees that the standard
counterterm prescription for AdS spacetimes yields a finite action
for the solutions.

The geometry can be foliated with surfaces of constant $r$, that we
call $\mathcal{B}_R$. These are defined by $f(r)=r-R=0$ and have
unit normal $n_\mu=\partial_\mu f/|\partial f|$. The induced metric
on these surfaces, $\gamma_{\mu\nu}$, and their extrinsic curvature
$K_{\mu\nu}$ are then
 \beq
\gamma_{\mu\nu}=g_{\mu\nu}-n_\mu n_\nu\,,\qquad
K_{\mu\nu}=\gamma_\mu^{\:\:\rho}\gamma_\nu^{\:\:\sigma}\nabla_\rho n_\sigma \,.
\label{InducedMetric}
 \eeq
The boundary
$\mathcal{B}$ of the lukewarm geometries is obtained by taking the
hypersurface $\mathcal{B}_R$ to infinity, \ie $R\rightarrow \infty$.
The boundary metric $h_{ab}$, where Roman indices run over the boundary coordinates, is then defined by taking the inverse of $\gamma^{ab}$, which in our case is simply $\gamma_{ab}$, and rescaling it by the inverse power of its conformal
factor, \ie
 \beq
h_{ab}=\lim_{{\:{\scriptstyle{\scriptstyle r\rightarrow
\infty}\:}}}\frac{\mathcal{A}^2}{r^2}\, \gamma_{ab}\qquad
\Rightarrow \qquad ds_{\sf  bdy}^2=-\mathcal{A}^2\ell^2 F(\chi) dt^2
+\frac{d\chi^2}{\mathcal{A}^2\ell^2
 F(\chi)G(\chi)}+G(\chi)d\phi^2\,. \label{boundaryMetric}
 \eeq
Close to the boundary $\chi$ tends to $x$, hence its range is
$x_0\leq\chi\leq x_B$ (we are in quadrants $II$ and $IV$). This
coordinate plays the role of radial coordinate of the boundary
metric. The resulting boundary geometry has some intriguing properties.
First notice that it is not a solution of $d=3$ Einstein-Maxwell
theory which indicates that there is an effective stress tensor,
$8\pi\GN_3 T_{ab}^{\rm eff}=
\mathcal{R}_{ab}-\frac{1}{2}\mathcal{R}h_{ab}$ (with
$\mathcal{R}_{ab}$ and $\mathcal{R}$ being the Ricci tensor and scalar
associated with $h_{ab}$) living on the boundary $\mathcal{B}$. This
is no surprise, since there is no dynamical gravity on the boundary: the conformal metric $h_{ab}$ describing the background on which
the dual CFT fields propagate simply fixes (part of) the boundary
conditions. Its geometry describes a boundary black hole with
horizons at the zeros of $F(\chi)$. The inner horizon is at
$\chi=y_3$ and the outer horizon is at $\chi=-b$. An investigation
of the large radial behaviour of \eqref{boundaryMetric} confirms
that it indeed describes a black hole with a well defined asymptotic
region. This exercise will also allow us to finally understand why we
choose only the lukewarm solutions that have a degenerate root of
$G(x)$ at $x=x_0$ \cite{Hubeny:2009ru}. Start by noticing that the
proper distance between any two points at radial coordinates $\chi_1$ and $\chi_2=x_0$ is given by
 \beq
\Delta \chi=\frac{1}{\mathcal{A}\ell}\int_{\chi_1}^{x_0}
\frac{d\chi'}{\sqrt{ F(\chi')G(\chi')}}\,.\label{ProperDistance}
 \eeq
From the above expression it is clear that $\Delta\chi$ diverges if and only if $x = x_0$ is a double root of $G(x)$. This point is then at an infinite proper spatial distance from any other point with $\chi\neq x_0$ and we naturally
identify $\chi=x_0$ as the asymptotic region of the boundary
geometry. Moreover, in its neighborhood, the norm of the stationary
Killing vector satisfies $\|\partial_t\|\rightarrow -1$, consistent
with the proposed interpretation. In turn, the spatial part of
\eqref{boundaryMetric} reduces to
 \beq
 ds_{2}^2=\frac{d\chi^2}{(\chi-x_0)^2}+(\chi-x_0)^2 d\phi^2\,. \label{AsympBoundaryMetric}
 \eeq
This is the metric on a Euclidean hyperboloid $\mathbb{H}^2$.
Therefore, the boundary black hole asymptotes to $\mathbb{R}\times
\mathbb{H}^2$.

\begin{figure}[ht]
\centerline{\includegraphics[width=.30\textwidth]{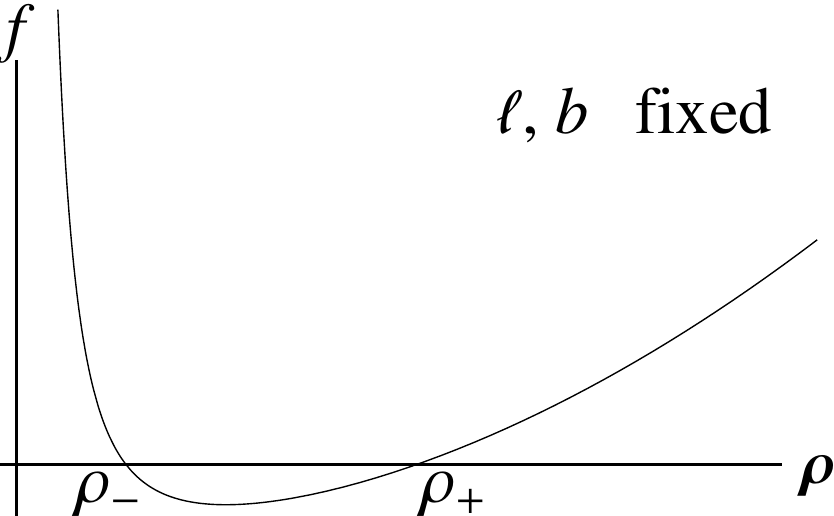}
\hspace{2cm}\includegraphics[width=.30\textwidth]{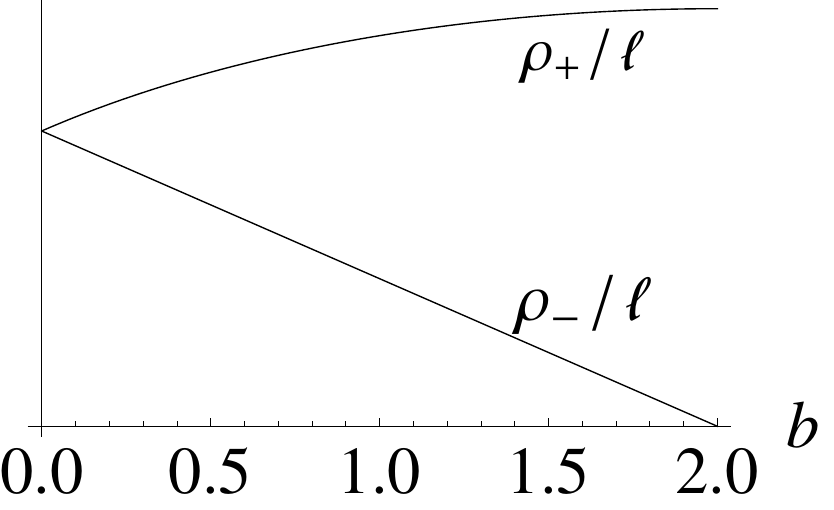}}
\caption{In the neighbourhood of $x=x_0$ the bulk lukewarm AdS
C-metric is described by \eqref{ZoomCmetric}. It represents a
charged hyperbolic black hole. The function $f(\rho)$ is represented
in the left plot, and has two hyperbolic horizons at $\rho_-$ and $\rho_+$.
Their location as a function of the parameter $b$ is represented in
right plot. For $b=0$, we have an extreme solution.}
\label{fig:PlanarBH}
\end{figure}
It is useful to look into the bulk charged AdS C-metric in the
neighborhood of the region $\chi \sim x_0$. We can zoom the bulk
metric on this asymptotic region by taking an appropriate scaling
limit. Concretely, take \eqref{AdSCmetric} with
\eqref{lukeDoubleProperties} and perform the scaling transformation
\cite{Hubeny:2009ru}
 \beq
 \epsilon \rightarrow 0\,,\qquad \hbox{with\; $t$,\; $y$,\; $\displaystyle X=\frac{x-x_0}{\epsilon}$\; and\; $\displaystyle \Phi=\frac{1}{2}\epsilon\phi$\; fixed}
 \,, \label{scaling}
 \eeq
which yields ($\mathcal{A}^2 \ell^2$ is given by
\eqref{lukeDoubleProperties})
 \eqn
 && A=q_e \,\ell\, y\, dt +\frac{\sqrt{1-16q_e^2}}{2\mathcal{A}}\,X \,d\Phi\,, \nonumber\\
 &&  ds^2=\frac{1}{\mathcal{A}^2(x_0-y)^2} \lp -
\mathcal{A}^2 \ell^2 F(y) dt^2
+\frac{dy^2}{F(y)}+\frac{2}{X^2}\,dX^2+2X^2d\Phi^2\rp.
\label{ScaleMetric}
 \eeqn
 Take now the coordinate transformation,
 \beq
t=\frac{\sqrt{2}}{\ell}\, T\,, \qquad
y=x_0-\frac{\sqrt{2}}{\mathcal{A} \rho}\,, \label{ScaleMetricAUX}
 \eeq
to get
 \eqn
&& A=\lp \sqrt{2}\, x_0-\frac{2}{\mathcal{A} \rho} \rp q_e\, dT +\frac{\sqrt{1-16q_e^2}}{2\mathcal{A}}\,X \,d\Phi\,, \nonumber\\
&& ds^2= -f(\rho) dT^2 +\frac{d\rho^2}{f(\rho)}+\rho^2 \lp
\frac{dX^2}{X^2}+ X^2d\Phi^2 \rp , \qquad
f(\rho)=\frac{\rho^2}{\ell^2}-1+\frac{(2\mathcal{A})^{-2}}{\rho^2}\,,
\label{ZoomCmetric}
 \eeqn
with $\mathcal{A}$ given again by \eqref{lukeDoubleProperties}. Notice
that the purely magnetic solution is obtained when one sets $q_e=0$
and one then has $q_m=1/4$, while the purely electric solution
corresponds to set $q_e=1/4$ and thus $q_m=0$. This is a solution of
AdS$-$Einstein-Maxwell theory \cite{Mann:1996gj}.
It describes a charged black hole with hyperbolic horizons, \ie with
$\kappa=-1$. Recall that the most general family of such a black
hole is given by
$f(\rho)=\frac{\rho^2}{\ell^2}-1+\frac{M}{\rho}+\frac{Q^2}{\rho^2}$.
So for our solution \eqref{ZoomCmetric} one has $M=0$ and
$Q=(2\mathcal{A})^{-1}$. Although the mass parameter vanishes in
this case, the geometry still describes a regular black hole with
inner $\rho_-$ and outer $\rho_+$ horizons clothing the curvature
singularity, because it is now in the $\kappa=-1$ class. These
horizons are located at $\rho_\pm=\sqrt{4\ell^2(1\pm b)\mp \ell^2
b^2}\,/(2\sqrt{2})$ (see Fig. \ref{fig:PlanarBH}).

\vskip 0.3cm
 At this point we can finally discuss in detail the
announced interpretation for our special family of lukewarm
C-metrics. They can be interpreted as black droplets and/or black
funnels in thermal equilibrium described by \eqref{AdSCmetric},
\eqref{AdSCmetricMaxwell} and \eqref{lukeDoubleProperties}. We can
rewrite these solutions in terms of the Fefferman-Graham coordinates
$(r,\chi)$ defined in \eqref{FGcoords}. When the radial coordinate
$r$ goes to infinity, one reaches the holographic boundary of the
droplets and funnels. At this boundary the droplets and funnels
induce a boundary black hole described by \eqref{boundaryMetric}.
The coordinate $\chi$ is a polar angle in the bulk, but in the
boundary it plays the role of a radial coordinate. The asymptotic
region of the boundary black hole is at $\chi=x_0$, and is at an
infinite spatial proper distance from any other point in the bulk by
\eqref{ProperDistance}. The black droplets and funnels live in range
$x_0\leq x\leq x_B$.
%
What remains are two quadrants $II$ and $IV$, each of which fully
describes a well-behaved solution (Fig.~\ref{fig:DiagDropFun}). The
$\partial_\phi$ axis of rotation at $x=x_B$ for the solutions in
quadrants $II$ and $IV$ is free of conical singularities if we
choose the period of $\phi$ to be \eqref{PeriodPhiFinal}. The axis
$x=x_0$ is also regular and describes points that are at arbitrarily
large distances from any other $\chi$. Quadrant $II$ is defined by
$y-x=1/r>0$ while quadrant $IV$ has $x-y=1/r>0$. The asymptotic
holographic boundary is shared by these two quadrants, and is at
$x-y=0$.

The solution of quadrant $II$ describes a black funnel. Indeed,
starting at point between $E$ and $A$ on the holographic boundary,
we are in the exterior of the boundary black hole with an outer
horizon represented by point $E$. As we enter into the bulk (towards
smaller values of $r=(y-x)^{-1}$), we will first cross the outer
horizon, $y=-b$, and then the inner horizon, $y=y_3$, of the black
funnel, until we hit the curvature singularity at $y=+\infty$
($r=0$). If we start at the holographic boundary already inside the
inner horizon of the boundary black hole, \ie at a point between $M$
and $N$, then as we move into the bulk we will remain inside the
bulk black funnel until we reach the curvature singularity at
$y=+\infty$. Therefore, this solution in quadrant $II$ indeed
describes a black funnel as sketched in Fig.
\ref{fig:dropletfunnel}.a. In that figure, we just represent the
outer horizon of the black funnel. In the bulk and for `large'
values of $\chi$ in the neighborhood of $\chi\sim x_0$ (the
`shoulder region'), the funnel approaches the geometry of a
hyperbolic black hole described by \eqref{ZoomCmetric}. At the
holographic boundary, the horizons of the black funnel match the
horizons of the boundary black hole.

On the other hand, the solution of quadrant $IV$ describes a black
droplet,  hovering over a planar black brane. Let us start at a
point between $A$ and $E$ at the holographic boundary, \ie between
the asymptotic region (point $A$) and outside the outer horizon
(point $E$) of the boundary black hole. We can dive into the bulk
keeping away from the $DE$ horizon until we first cross the outer
horizon $y=b-4$, then the inner horizon $y=y_0$, and finally hit the
curvature singularity $y=-\infty$ ($r=0$) of the deformed hyperbolic
black hole. Again, for `large' values of $\chi$ in the neighborhood
of $\chi\sim x_0$, this deformed hyperbolic black hole approaches
the geometry of a hyperbolic black hole described by
\eqref{ZoomCmetric}. Instead, we can start inside the inner horizon
of the boundary black hole, \ie in between points $M$ and $N$. As we
enter into the bulk, we will at first stay inside the inner horizon
($y=y_3$) of the black droplet. As we keep going deeper into the
bulk, we will cross the outer horizon ($y=-b$) of the black droplet.
These horizons of the black droplet coincide with the horizons of
the boundary black hole at the holographic boundary. At this stage
we are outside any black hole horizon but, as we keep moving deeper
into the bulk, we will cross the outer horizon $y=b-4$, then the
inner horizon $y=y_0$, and finally we will hit the curvature
singularity $y=-\infty$ of the deformed hyperbolic black hole. This
description of the solution in quadrant $IV$ indeed corresponds to
the black droplet interpretation that is sketched in Fig.
\ref{fig:dropletfunnel}.b. In this figure, we again just represent
the outer horizon of the black holes of the system. It is important
to notice that the black droplet of quadrant $IV$ and the black
funnel of quadrant $II$ share the same boundary, and that their
outer horizons have the same temperature.

\vskip 0.3cm
 To finish this section we address the issue of the
acceleration parameter source. In the standard interpretation of the
C-metric as a pair of accelerated black holes
\cite{Plebanski:1976gy}, one always works in a range of coordinates
where we choose the period of $\phi$ to eliminate a conical
singularity at one of the poles. This leaves a singularity on the
other pole of $\partial_\phi$ that is interpreted as a cosmic string
or strut that sources the acceleration $\mathcal{A}$ of the black
holes, as explained in the previous subsection. However, if we work instead
in the subsector of C-metric solutions where $G(x)$ has a double
root at $x=x_0$, this point is infinitely far apart from any other
point $x$ in $x_0\leq x\leq x_B$. We can then choose the
period of $\phi$ that eliminates the conical singularity at $x=x_B$;
\eg for the lukewarm solutions this period is given by
\eqref{PeriodPhiFinal}. The potential deficit angle \eqref{deficit} at $x=x_0$
vanishes because $G'(x_0)=0$ when $x_0$ is a double root: $x=x_0$ is
at asymptotic infinity and the solution is then also regular here
without any conical singularity/cosmic string. The fact that these
solutions with a double root of $G(x)$ are regular everywhere is
satisfying but the lack of a physical source that explains the
origin of the acceleration parameter is puzzling. Futhermore, it is
not clear how we can measure the acceleration parameter that
characterizes these solutions. Indeed, although the mass and the
charges of the solution are conserved charges that can be measured
at infinity, we seem to lack an explicit measure of $\mathcal{A}$.
Notice that this issue applies equally well to the uncharged
solutions studied in \cite{Hubeny:2009ru,Hubeny:2009kz}. It is
important to try to understand this issue.

We think that the appropriate explanation borrows some ideas learned
with a completely different solution, namely the dipole black rings \cite{Emparan:2004wy}.
In short, and in the simplest case, these are black ring solutions
of supergravity that are electrically coupled to a two-form
potential (or to a dual magnetic one-form in five dimensions) \ie
they create a field analogous to a dipole. The keypoint for our
purposes is that these solutions are characterized by more
parameters than conserved charges (which can be measured at
infinity). In the simplest case, the five-dimensional dipole black
ring is characterized by three parameters, but only has two
conserved charges (mass and angular momentum). The third parameter
is a non-conserved dipole charge that cannot be  measured at
infinity. Instead, to compute it we have to do a {\it local}
operation where we consider an $S^2$ that encloses a section of
the ring (that is locally a string). The dipole charge is then given
by the electric field flux across this $S^2$.

The droplet/funnel system seems to have some resemblance with the
dipole rings in the sense that it is also a solution that is
characterized by more parameters than conserved charges. The
acceleration parameter $\mathcal{A}$ is not a conserved charge.
Instead, to determine it we also need to do a kind of a local
measurement as follows. We already saw that in the neighborhood of
$\chi \sim x_0$ the black droplets and black funnels share the same
geometry, namely \eqref{ZoomCmetric}. In this region the geometry
describes a hyperbolic black hole characterized only by its charge
that is proportional to the acceleration, $Q=(2\mathcal{A})^{-1}$.
To measure $\mathcal{A}$, we can then make a local measurement of
the decay of its associated electromagnetic field in the asymptotic
region $r\rightarrow \infty$ and $\chi\rightarrow x_0$.

\setcounter{equation}{0}
\section{Free energy of the black droplets and funnels
\label{sec:FreeEtensor}}

The black funnel of quadrant $II$ and the black droplet of quadrant
$IV$ share  the same holographic boundary, where they induce the
same boundary metric. Furthermore, they are in thermal equilibrium
since their outer horizons have the same temperature. Therefore,
given the boundary data, both geometries contribute to the
gravitational partition funtion. In this section, we shall compute
the Euclidean action of the black funnel and the black droplet, thus
showing which of the configurations dominates the thermal ensemble.
We shall work only with the canonical ensemble for the magnetic
solutions, for the reasons we will now detail.

Let us address the issue of regularity of the electromagnetic field
for  the black droplet. The charge and the associated chemical
potential are relevant for the thermodynamic description of the
system. Let us imagine that we had rotation instead of charge in our
solutions. If the angular velocities of the two event horizons did
not match, the Euclideanised solution would possess a conical
singularity, as it would if the temperatures did not match. The
charged case is different though, since the associated
irregularities are not in the geometry itself. Nevertheless, the
chemical potential in our black droplet solutions cannot be made to
match between the black droplet and the deformed brane horizons.
Since the two horizons cannot be in equibrium in this sense, we
shall not work with the grand-canonical ensemble.

Consider the electromagnetic potential $A$
\eqref{AdSCmetricMaxwell}.  Let us assume first that $q_e=0$, i.e.
we have a magnetic solution. In the well known case of a spherical
magnetic monopole, $A$ is required to be regular (e.g. $A^2$ finite)
on each segment of the axis $\theta=0,\pi$. One can consider two
patches, each covering one segment, on which a gauge is chosen such
that $A$ vanishes on the axis. Then the consistency condition for
gluing the two patches is that particle charges on the monopole
background are quantised. In our case, we have a single segment of
the rotation axis, $x=x_B$, but we also have the spatial limit
$x=x_0$ on which $A^2$ diverges, unless a convenient gauge is
chosen. Therefore, we also need two patches,
\begin{equation}
x=x_B \,: \quad A \to A - d \left( \frac{q_m}{\mathcal{A}}x_B \phi \right) \,, \qquad
x=x_0 \,: \quad A \to A - d \left( \frac{q_m}{\mathcal{A}}x_0 \phi \right) \,.
\end{equation}
whose gluing leads to a quantisation condition on particle charges $e$,
\begin{equation}
e \,\frac{q_m}{\mathcal{A}} \,(x_B-x_0)= 2\pi n,
\end{equation}
where $n$ is an integer.

Consider now the electrically charged solutions. As we mentioned,
there is  a difference in the electric potential $\phi_E=-q_e \ell
y$ in \eqref{AdSCmetricMaxwell} between the two horizons, located at
$y=-b$ and $y=b-4$. The corresponding electric field would lead to a
current between the horizons if any charged particles were present.
(There is always at least one type of charged particle in
Einstein-Maxwell theory, which is a mini black hole produced as a
quantum fluctuation, but this effect is suppressed as $1/N$ in the
AdS/CFT correspondence.) Notice that the usual quantisation
procedure, as discussed above for the magnetic case, fails here. One
could consider two patches, such that each covers one horizon, and
on each patch there is a simple gauge choice which makes $A_\mu$
regular at that horizon.  However, that gauge transformation would
depend on the time coordinate. Since, in the Euclideanised
spacetime, we have $t \to -i \tau$, we cannot make sense of the
corresponding gauge transformations for charged particles in the
system. Since we cannot find a meaningful way of working with the
electric solutions, even in the canonical ensemble, we shall
consider magnetic solutions only, \ie $q_e=0$.

We now proceed to the computation of the free energy of the magnetic
black funnel and black droplet. We will be comparing these solutions
at the same values of the temperature and of the magnetic charge,
i.e. we consider here the canonical ensemble. The relevant
thermodynamic potential is therefore the Helmholtz free energy.

The Helmholtz free energy $\mathcal{F}$ of a solution is given by the
ratio of the Euclidean action of the solution $I$ and its thermal
period $\beta$, $\mathcal{F}=I/\beta$. The Euclidean action includes
the AdS-Einstein-Maxwell integral over the bulk $\cal M$, $I_{\rm
bulk}[g_{\mu\nu},A_\mu]$, plus the York-Gibbons-Hawking surface integral
over the boundary $\partial\cal M$, $I_{\rm YGH}[g_{\mu\nu}]$. The
latter is required so that upon variation with metric fixed at the
boundary, the action yields Einstein's equations. This is not the
whole story. Since we are in an asymptotically AdS background, the
sum of these contributions, $I_{\rm bulk}+I_{\rm YGH}$,
diverges because the volumes of both $\cal M$ and $\partial\cal M$
are infinite (and the integrands are nonzero). The correct and finite
computation of the gravitational action then requires regulating the
action. In the counterterm subtraction approach, the divergences are
eliminated by adding a counterterm contribution to the action,
$I_{\rm ct}[\gamma_{ab}]$ \cite{sken,pervijay,Emparan:1999pm}.
This is an extra surface integral that depends only on the induced boundary metric
$\gamma_{ab}$ and its derivatives. If we work in Fefferman-Graham coordinates, the
expression for $I_{\rm ct}[\gamma_{ab}]$ is universal, depending only
on $\ell$ and the spacetime dimension.

The finite Euclidean gravitational action for a solution of
AdS-Einstein-Maxwell theory, and including only the counterterms
that make a contribution in  four dimensions, is then given by
\cite{sken,pervijay,Emparan:1999pm}
\begin{equation}
I=I_{\rm bulk}[g_{\mu\nu},A_\mu]+I_{\rm YGH}[g_{\mu\nu}] +I_{\rm ct}[\gamma_{ab}]\,, \label{totalAction}
\end{equation}
with \eqn && I_{\rm bulk}=-\frac1{16\pi\GN}\int_{\cal M}
d^4x\,\sqrt{g} \left(R+\frac{6}{\ell^2} - F_{\alpha\beta}F^{\alpha\beta}\right)\,,\qquad I_{\rm
YGH}= -\frac1{8\pi\GN}\int_{\partial\cal M} d^3x\,
\sqrt{\gamma} K\,,\nonumber\\
 &&
 I_{\rm ct}=\frac1{8\pi\GN}\int_{\partial\cal M}
d^3x\,\sqrt{\gamma} \lp\frac2{\ell}+\frac{\ell}{2}{\cal R} \rp .
\label{ActionContrib}
 \eeqn
Here, $R$ is the Ricci scalar of the bulk Euclidean metric
$g_{\mu\nu}$, ${\cal R}$ is the Ricci scalar of the induced boundary
metric $\gamma_{ab}$, and $K=\gamma^{ab} K_{ab}$ is the trace of the
extrinsic curvature of the boundary $\partial\cal M$ as embedded in
$\cal M$ defined in \eqref{InducedMetric}.\footnote{ For a non-zero
electric charge, the Hawking-Ross surface integral, $I_{\rm
HR}[g_{\mu\nu},A_\mu]$, also contributes to the total action in the
canonical ensemble \cite{Hawking:1995ap}. This term allows us to
impose fixed electric charge $q_e$ as a boundary condition at
infinity, and is required by the electromagnetic duality. Denoting
with a  superscript $(e)$ the electric sector of the electromagnetic
field, the total Euclidean action is given by $I=I_{\rm bulk}+I_{\rm
YGH}+I_{\rm HR} +I_{\rm ct}$, with
 \eqn
 I_{\rm HR}=-\frac1{4\pi\GN}\int_{\partial\cal M} d^3x\,\sqrt{\gamma}
 F^{(e)\mu\nu}A^{(e)}_\mu n_\nu \,.
\label{ActionContribElectric}
 \eeqn}

A word of caution is in order here. It is widely assumed in the
literature that the expression for $I_{\rm ct}$ as written in
\eqref{ActionContrib} provides a finite {\it covariant} definition
of the gravitational action for asymptotically AdS spaces. That is,
it is often assumed that one may use this expression for
$I_{\rm ct}$ to regulate the action for {\it any} choice of
coordinates on any asymptotically AdS solution. However, this is
not quite correct \cite{Papadimitriou:2005ii}. Indeed the standard
counterterms as written in \eqref{ActionContrib} are derived using a
Fefferman-Graham coordinate system. These counterterms, defined on
hypersurfaces of constant Fefferman-Graham (FG) radial coordinate,
are not necessarily the correct counterterms on the hypersurfaces of
constant radial coordinate in all other coordinate systems. If we choose instead to
work in a coordinate system where the radial coordinate does not
agree with the radial FG coordinate we should redo the asymptotic
analysis for the new radial coordinate from the very beginning using
a regulator for this new radial coordinate and rederive the
appropriate counterterms.

This is an important issue in the current analysis because we find
that if we work with the C-metric in the $\{t,r,x,\phi\}$ coordinate
system, \ie with $y=x\pm 1/r$ and $x=\chi$ instead of
\eqref{FGcoords}, then the total action with the counterterm as written in
\eqref{ActionContrib} does not yield a finite result. This is a particular case where the expression
\eqref{ActionContrib} derived using a FG coordinate system does not
hold on another coordinate system. Another example where a similar
situation occurs is the Kerr-Newman black hole: if we work in the
standard Boyer-Lindquist coordinate system, expression
\eqref{ActionContrib} for the counterterm action also fails to
render the total action finite \cite{Papadimitriou:2005ii}.

We choose to work with the well established FG counterterm action
\eqref{ActionContrib}. The appropriate perturbative definition of
the FG coordinate system is that, up to sufficient order in $1/r$,
the coefficient of the radial line element should not depend on the
polar angle, and there cannot be a mixed term between the FG radial
$r$ and angular the $\chi$ coordinate. The criteria to decide which
coefficients of this series expansion in $1/r$ vanish is the
requirement that the gravitational action \eqref{totalAction} is
finite. With this prescription, we find that the Fefferman-Graham
coordinates for the C-metric are those written in \eqref{FGcoords}.
The coordinate transformation in the polar angle $x(\chi,r)$ was
chosen to guarantee that the cross term between the radial and polar
component of the metric vanishes at infinity up to order $1/r^2$,
\ie $g_{ri}\sim \mathcal{O}(1/r^2)$ as $r\rightarrow\infty$ (with
$i$ being a spacelike boundary coordinate). This is enough to obtain
a finite action as we confirm next.

\begin{figure}[ht]
\centerline{\includegraphics[width=.40\textwidth]{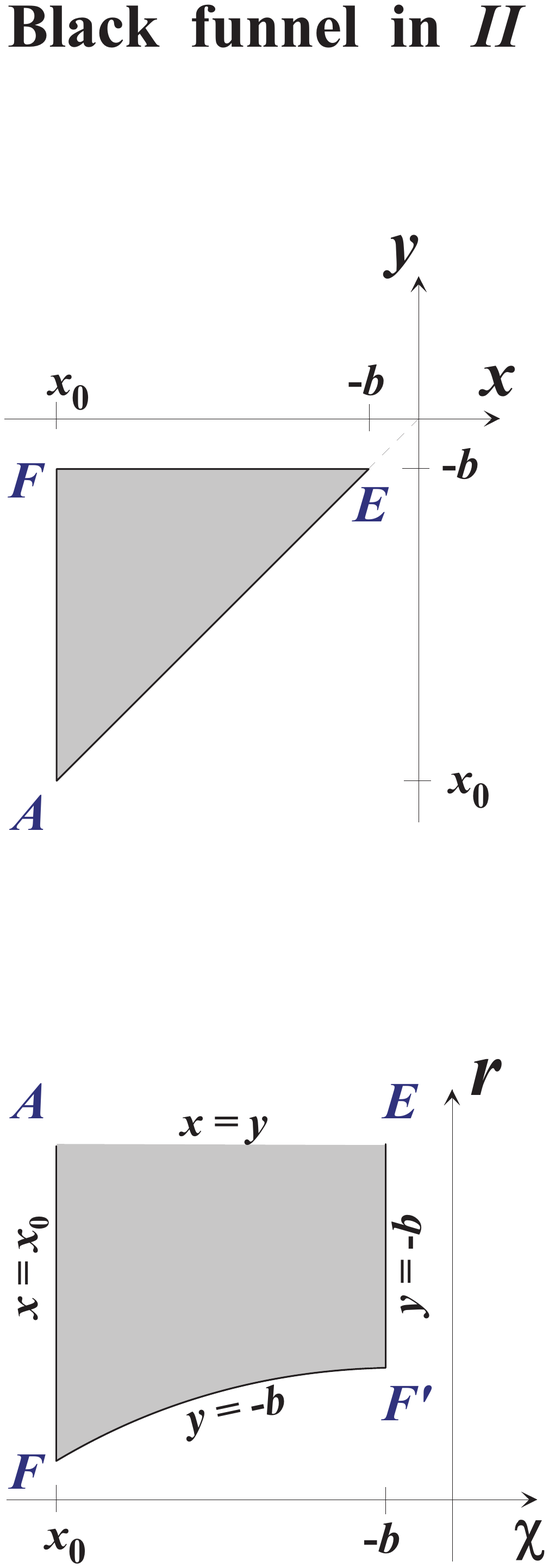}
\hspace{2cm}\includegraphics[width=.40\textwidth]{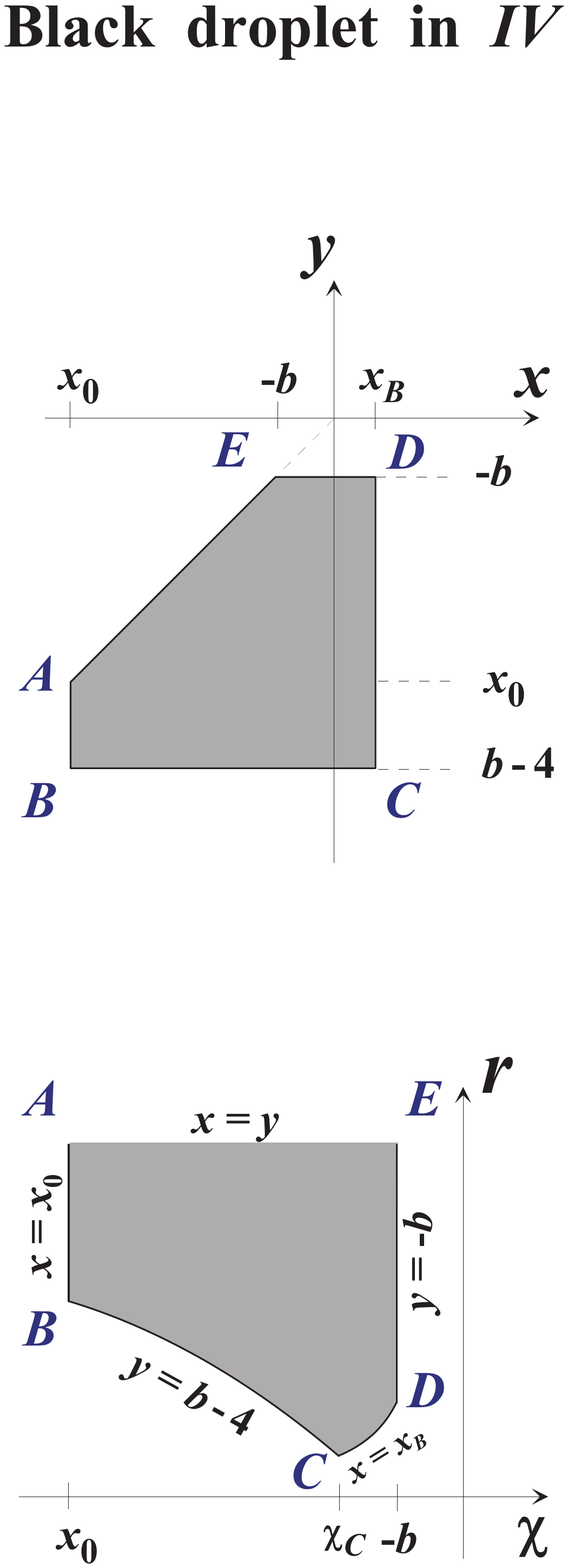}}
\caption{Map between the boundaries of the black funnel and black
droplet in the original coordinates $(x,y)$ and the $(\chi,r)$
Fefferman-Graham coordinates defined in \eqref{FGcoords}. The left
panel describes the black funnel of quadrant $II$, whose single
horizon is  split into two branches in the $(\chi,r)$ coordinates:
$FF'$ and $F'E$. This solution shares its asymptotic holographic
boundary $AE$ with the black droplet of quadrant $IV$ described in
the right panel. The black droplet and black brane horizons
correspond to the $DE$ and $BC$ curves, respectively, and are
disjoint.} \label{fig:CoordDropFun}
\end{figure}

To evaluate the bulk and surface integrals in \eqref{ActionContrib}
we need to define the position of the horizons, of the
boundary and of the azimuthal axis in the Fefferman-Graham
coordinates \eqref{FGcoords}. The map between the black
droplet/funnel regions in the $(x,y)$ coordinates and the $(\chi,r)$
FG coordinates is represented in Fig. \ref{fig:CoordDropFun}.

The left panel of Fig. \ref{fig:CoordDropFun} shows this map for the
black funnel of quadrant $II$. The closed curve [$AEF$] is in
correspondence with the associated curve in quadrant $II$ of Fig.
\ref{fig:DiagDropFun}. Particular care must be taken when mapping
the segment $[EF]$, representing the horizon, to $(r,\chi)$
coordinates, since its intersection point with asymptotic infinity
$F$ is resolved by the map \eqref{FGcoords} into a curve $[FF']$. To
see this, we cut the spacetime at a coordinate distance
$\varepsilon>0$ from the horizon, so that the triangle in the upper
Fig.~\ref{fig:CoordDropFun}.a ends on a segment $[\hat E,\hat F]$ at
$y=-b-\varepsilon$. This segment is mapped to a smooth curve in the
FG coordinate plane, but as the regulator is removed in the limit
$\epsilon\rightarrow0$, this curve becomes non-differentiable at the
point $F'$. This breaks the full curve $[EF]$ into the curve $[FF']$
described by $r(\chi)=-\frac{\mathcal{A}^2\ell^2 F(\chi)}{\chi+b}$,
and the vertical line $[F'E]$ at $\chi=-b$. Since the full curve
describes the boundary of the integration region and it remains
piecewise differentiable, the appearance of this singularity is
harmless and the integration can be performed directly over the
$[AFF'E]$ region of Fig.~\ref{fig:CoordDropFun}.a in FG coordinates.
This black funnel shares its asymptotic boundary $[AE]$ with the
black droplet of quadrant $IV$ . The right panel of Fig.
\ref{fig:CoordDropFun} presents the map $(x,y)\rightarrow (\chi,r)$
for the latter black droplet. The closed curve [$ABCDE$] is in
correspondence with the associated curve in quadrant $IV$ of Fig.
\ref{fig:DiagDropFun}. The curve $BC$ is described by
$r(\chi)=\frac{\mathcal{A}^2\ell^2 F(\chi)}{\chi-b+4}$. The curve
$CD$ is described by $r(\chi)=\frac{\mathcal{A}^2\ell^2
G(\chi)}{x_B-\chi}$. These two curves meet at point $C$, for which
$\chi=\chi_C$.

With this information, we can finally compute the integrals in
\eqref{ActionContrib}. The bulk integrals $\int_{\cal M}$ in $I_{\rm
bulk}$ for the black funnel in quadrant $II$ and the black droplet in
quadrant $IV$ have the following schematic structure,
 \eqn
 && \hspace{-0.5cm}\hbox{Black funnel in $II$:} \qquad \int_{\cal M}\rightarrow
\int_{0}^{\beta} d\tau \int_{0}^{\Delta\phi} d\phi \int_{x_0}^{-b}
d\chi \int_{-\frac{\mathcal{A}^2\ell^2
F(\chi)}{\chi+b}}^{\infty} dr ,\label{IntegralsIIandIV} \\
 && \hspace{-0.5cm} \hbox{Black droplet in $IV$:} \qquad \int_{\cal M}\rightarrow \int_{0}^{\beta} d\tau \int_{0}^{\Delta\phi} d\phi
\left[ \int_{x_0}^{\chi_C} d\chi \int_{-\frac{\mathcal{A}^2\ell^2
F(\chi)}{b-4-\chi}}^{\infty} dr +\int_{\chi_C}^{-b} d\chi
\int_{\frac{\mathcal{A}^2\ell^2 G(\chi)}{x_B-\chi}}^{\infty}
dr\right],\nonumber
 \eeqn
and the boundary integrals $\int_{\partial \cal M}$ in $I_{\rm YGH}$
and $I_{\rm ct}$ are given by the same integration structure but
with the radial integration removed. In these integrals, the thermal
period $\beta$ is given by \eqref{betaFinal} and the azimuthal
period $\Delta\phi$ is given by \eqref{PeriodPhiFinal}.

Evaluating these integrals we finally find that the total, finite,
action for the black droplets and black funnels is given by
\begin{eqnarray}
&\!\!\textrm{Black funnel in $II$:} \quad & I_{{\rm BF}}=\frac{\pi
(2-b)^3}{64\sqrt{2}\,\mathcal{G}}\,, \nonumber\\
&\textrm{Black droplet in $IV$:} \quad & I_{{\rm BD}}=\frac{\pi
(2-b)^2\lp b-2+8\sqrt{2}\rp}{64\sqrt{2}\,\mathcal{G}}\,.
\label{ActionFinal}
\end{eqnarray}

\begin{figure}[ht]
\centerline{\quad\includegraphics[width=.45\textwidth]{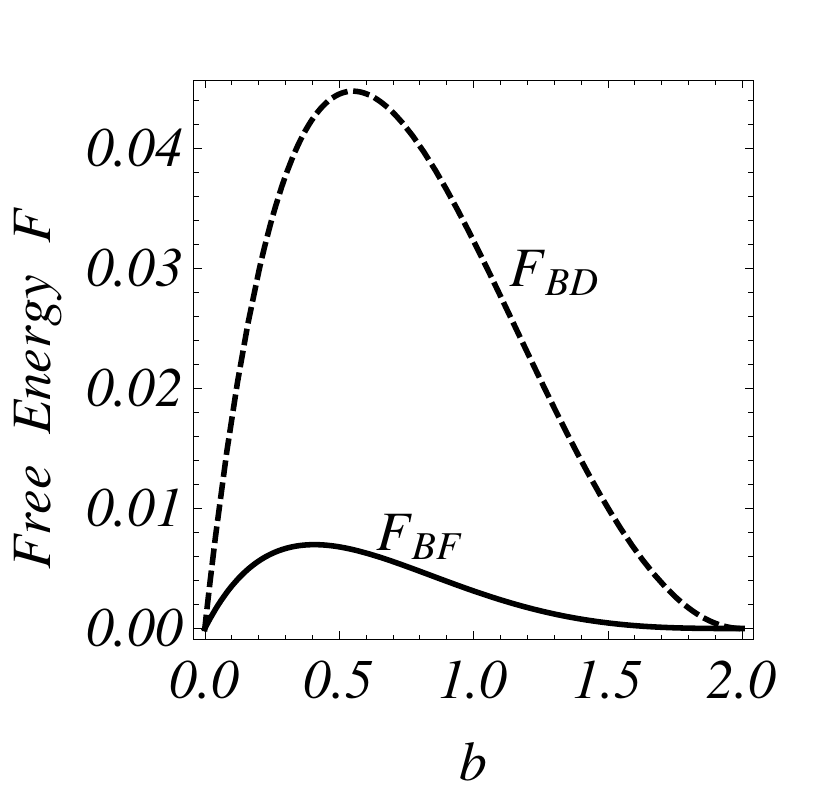}
\hfill\includegraphics[width=.43\textwidth]{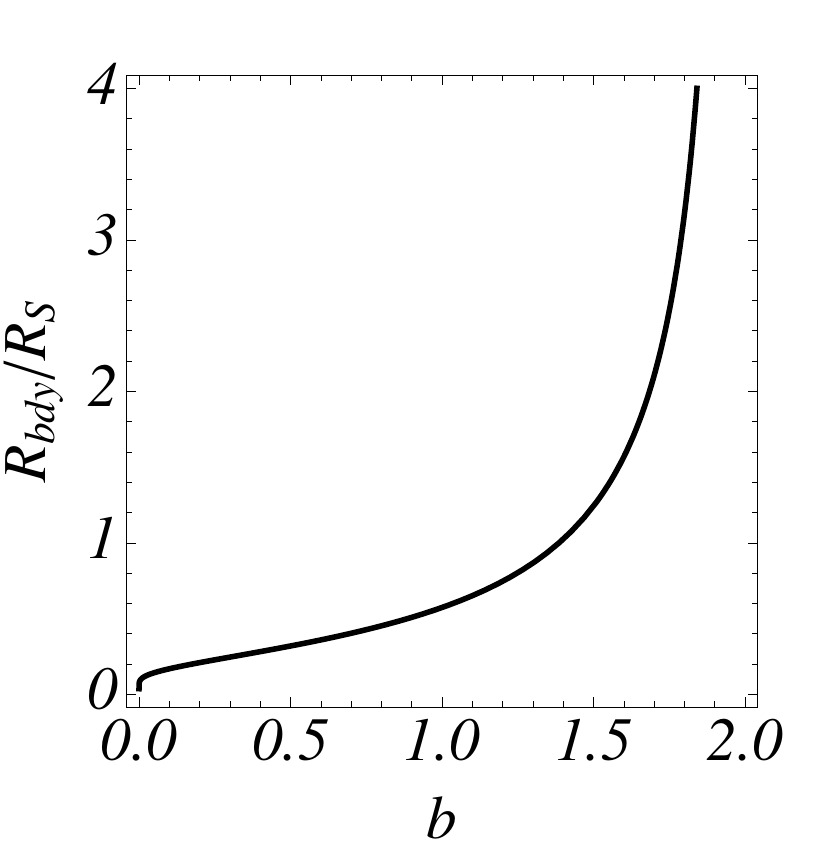}\quad}
\caption{{\it Left Panel}: Free energy for the black droplets ($BD$)
and black funnels ($BF$) as a function of the parameter $b$ for
fixed $\ell$ ($\mathcal{G}=1$). The free energies of both solutions
vanish at $b=0$ because the temperature vanishes there, while the
free energies are zero at $b=2$ because the associated Euclidean
actions vanish there. {\it Right Panel}: Size of the boundary black
hole, $R_{\sf bdy}$ {\it vs} the distance $R_{\sf S}$ between the
``shoulder" of the hyperbolic black brane or black funnel (see
Fig.~\eqref{fig:dropletfunnel}) and the holographic boundary.}
\label{fig:CompareFreeEnergy}
\end{figure}

We can know compare the Helmholtz free energy $\mathcal{F}=I/\beta$
of the pairs of black droplets/funnels that share the same
holographic boundary, charge and temperature. This is done in the
left panel of Fig.~\ref{fig:CompareFreeEnergy}. The black funnels
always have a smaller free energy than the black droplets and
therefore they always dominate the partition function. In
particular, there is no phase transition between these two
solutions, except possibly at $b = 2$ where the two free energies
vanish. However, at this $b=2$ point the two horizons merge, leaving
no static external spacetime region.

\begin{figure}[ht]
\centerline{\raisebox{1.5cm}{\includegraphics{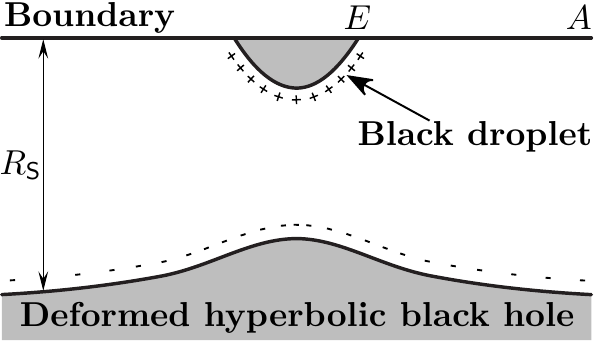}}
\hspace{1.5cm}
\includegraphics[width=.40\textwidth]{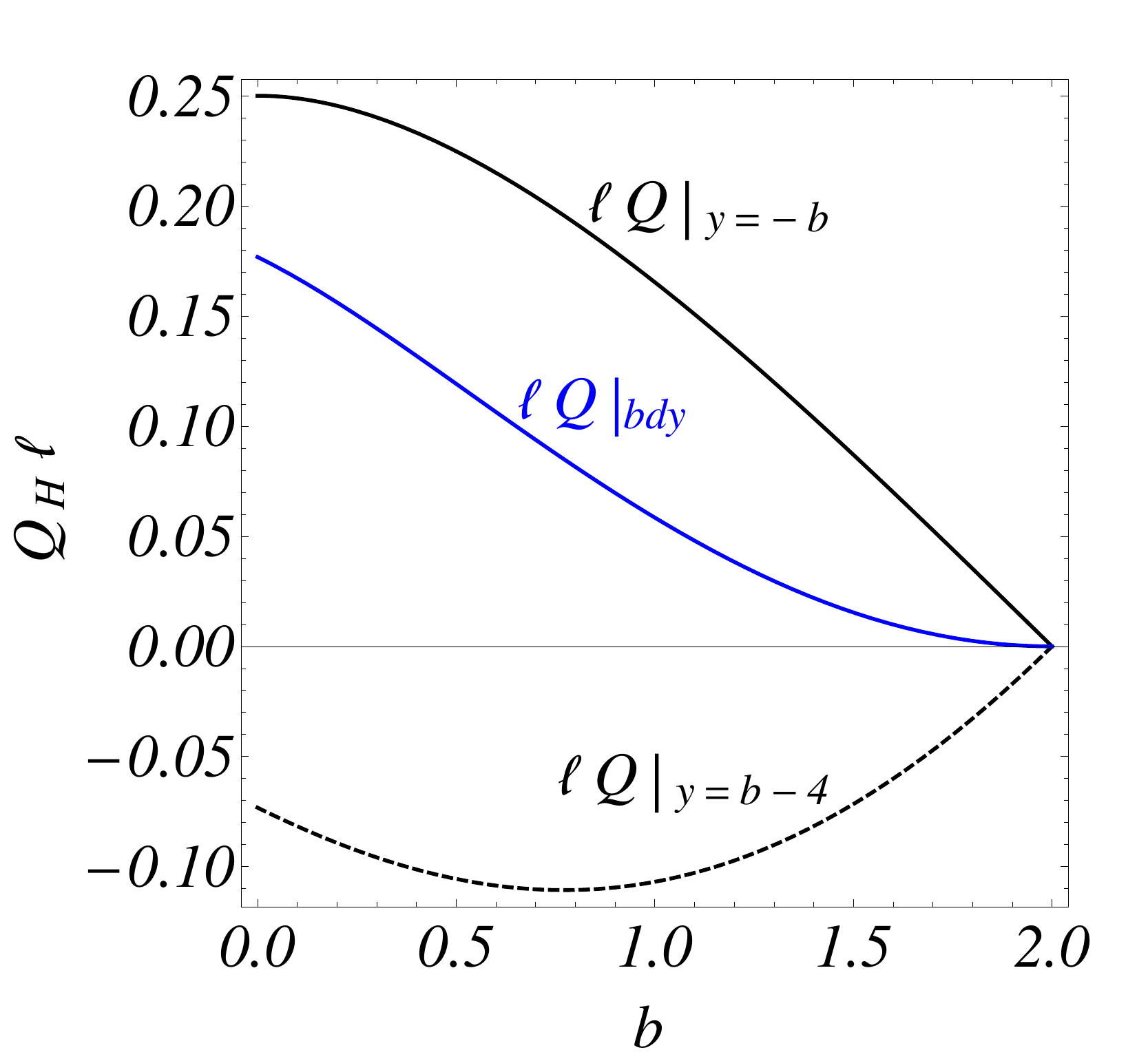}}
\caption{{\em Left panel:} The distance $R_{\sf S}$ between the ``shoulder'' region of the planar black hole and the boundary, and the distribution of charges on the horizons. {\em Right panel:} Charge of the black droplet horizon $\ell Q{\bigr
|}_{y=-b}$, charge of the deformed hyperbolic horizon $\ell Q{\bigr
|}_{y=b-4}$ and charge of the boundary black hole horizon $\ell
Q{\bigr |}_{\sf  bdy}$. The latter satisfies \eqref{Qdroplet} and
also gives the black funnel horizon charge.} 
\label{fig:charge}
\end{figure}

This conclusion is worth of some discussion. The notions of black
funnels and black droplets and their interpretation were first
introduced in \cite{Hubeny:2009ru}. Explicit examples of droplets
and funnels, constructed out of the (uncharged) AdS C-metric, were
also presented in \cite{Hubeny:2009ru,Hubeny:2009kz}. In these works
it was advocated that there should be a phase transition
between black funnels and droplets. For {\it asymptotically flat}
boundary black holes, the transition point is expected to be
determined by the ratio $R_{\sf bdy}/R_{\sf S}$. Here, $R_{\sf bdy}$
is the typical size of the boundary black hole, and $R_{\sf S}$ is
the distance between the ``shoulder" of the planar black brane or
black funnel and the holographic boundary, as shown in the left panel of Fig.~\ref{fig:charge} (in the neutral case this
distance is given by the inverse of the temperature $R_{\sf S}\sim 1/T_H$). Ref. \cite{Hubeny:2009ru} argues that one should
expect the black droplets to be dominant for $R_{\sf bdy}/R_{\sf
S}\ll 1$ and that the black funnels should dominate on the
complementary regime, $R_{\sf bdy} /R_{\sf S} \gg 1$. However, the
authors of \cite{Hubeny:2009ru,Hubeny:2009kz} did not test their
prediction against the results for the free energy of their explicit
droplets and funnels. The reason is that the black droplets
constructed out of their uncharged AdS C-metric can never be in
thermal equilibrium, \ie the droplet and deformed hyperbolic horizon
can never be at the same temperature when the charge vanishes. As we
explicitly found, this is no longer the case when we switch on the
Maxwell field: the magnetic lukewarm black droplets are in thermal
equilibrium. This allowed us to compare the free energies for the
magnetic lukewarm black droplets and funnels, and we found no sign of a phase
transition.
 Our boundary black holes asymptote to $\mathbb{R}\times
 \mathbb{H}^2$ (and are charged) instead of being asymptotically flat and thus
the phase transition estimative of
\cite{Hubeny:2009ru,Hubeny:2009kz} is {\it not} expected to hold. It
is nevertheless interesting to check what happens. We will do it
trying to follow as close as possible the arguments of
\cite{Hubeny:2009ru,Hubeny:2009kz}.

Consider first the boundary black hole size  $R_{\sf bdy}$. Since
our boundary black holes asymptote to $\mathbb{R}\times
\mathbb{H}^2$, circles of constant radial coordinate $\chi$ shrink
as $\chi$ increases. 
Therefore, we should not take, \eg the proper
size of constant $\chi$ circles as a measure of $R_{\sf bdy}$. On
the other hand, the volume outside the boundary horizon is finite
and decreases as the black hole size increases.\footnote{We thank
Don Marolf for this observation.} Therefore a good measure of
$R_{\sf bdy}$ seems to be the inverse of this volume, \ie
\begin{equation}
R_{\sf  bdy} \simeq \left[\int_0^{\sqrt{2}\,\pi}d\phi
\int_{x_0}^{-b}d\chi\, \sqrt{\sigma} \right]^{-1}\,, \label{Rbdy}
\end{equation}
where $\sigma$ is the determinant of the induced metric
$\sigma_{ab}=h_{ab}+N_aN_b$ on a timelike surface $f(t)=t-t_0=0$,
with normal $N_a=-\partial_a f/|\partial f|$. (See
\eqref{boundaryMetric}, \eqref{PeriodPhiFinal}, and recall that for
quadrants $II$ and $IV$ the outer horizon is at $\chi=-b$). To find
$R_{\sf S}$, \ie the distance between the ``shoulder" of the
hyperbolic black brane or black funnel (see
Fig.~\eqref{fig:dropletfunnel}) and the holographic boundary, recall
that the asymptotic ``shoulder" region is described by
\eqref{ZoomCmetric}. Introducing the new holographic coordinate
$z=-\ell^2/\rho$, the gravitational field of \eqref{ZoomCmetric}
reads
\begin{equation}
ds^2=\frac{\ell^2}{z^2}\left[-f(z)dT^2+\frac{dz^2}{f(z)}+\ell^2\left(\frac{dX^2}{X^2}+X^2
d\Phi^2\right)\right]\, \qquad \hbox{with} \quad
f(z)=1-\frac{z^2}{\ell^2}+\frac{1}{4\mathcal{A}^2\ell^2}\frac{z^4}{\ell^4}\,.
\label{RSmetric}
\end{equation}
The holographic asymptotic boundary is at $z=0$ and the (outer)
horizon position, $z=z_+\equiv R_{\sf S}$, determines the desired
distance between the holographic boundary and the brane/funnel
``shoulder"\footnote{One could refine the definition of $R_{\sf bdy}$ by using the inverse temperature of the hyperbolic brane \eqref{RSmetric} or the proper distance between the planar black brane and the black droplet instead of the holographic distance $z_+$. However,  this does not affect qualitatively our discussion.}:
\begin{equation}
R_{\sf S} =\sqrt{2} \mathcal{A}\ell\sqrt{1-\sqrt{
1-\frac{1}{\mathcal{A}^2\ell^2} }}\,. \label{RS}
\end{equation}
The ratio $R_{\sf  bdy}/R_{\sf S}$ for the families of funnels
and droplets that we analyzed is plotted in the right panel of
Fig.~\ref{fig:CompareFreeEnergy}. So, if the arguments of
\cite{Hubeny:2009ru,Hubeny:2009kz} were valid also for
asymptotically hyperbolic boundary black holes, we would expect the
black droplets to dominate a low values of $b$ ($R_{\sf bdy}\to 0$
as $b\to 0$). This is not the case, as the left panel of Fig.
\ref{fig:CompareFreeEnergy} shows. In sum, the outcome of our
exercise emphasizes that the phase transition estimative of
\cite{Hubeny:2009ru} does not hold for boundary black holes that
asymptote to $\mathbb{R}\times \mathbb{H}^2$, as pointed out in
\cite{Hubeny:2009kz}.

To have a better physical understanding of our results, note that
the charge of the system contributes to the fact that black funnels
dominate the partition function. A good strategy to confirm this
statement is to look at the charges of the black funnel and droplet
horizons, as defined by the pullback of the Maxwell field strength
on the horizon hypersurface. As before, we will consider the
magnetic charge only, i.e. $q_e=0$. In these conditions, the 2-form
associated with the Maxwell potential \eqref{AdSCmetricMaxwell},
written in terms of the Fefferman-Graham coordinates
\eqref{FGcoords}, is
\begin{equation}
F=\frac{q_m}{\mathcal{A}\,r^2}\left[ \varepsilon \mathcal{A}^2
\ell^2 G(\chi) dr\wedge d\phi + r
\left(r-\varepsilon\,\mathcal{A}^2 \ell^2
G'(\chi)\right)d\chi\wedge d\phi \right]\,, \label{Max2form}
\end{equation}
where $\varepsilon=1$ in quadrant $II$, $\varepsilon=-1$ in quadrant
$IV$; see \eqref{FGcoords}.

We can consider the pullback of $F$ to a hypersurface $r=R(\chi)$,
which yields
\begin{eqnarray}
F{\bigr|}_{r=R(\chi)}=\rho_R \, d\chi\wedge d\phi\,,\qquad
\hbox{where}\quad
\rho_R = \frac{q_m}{\mathcal{A}}\left[1-\varepsilon
\mathcal{A}^2 \ell^2\lp\frac{G(\chi)}{R(\chi)}\rp'\,
\right]
\label{MaxPullBack1}
\end{eqnarray}
is the charge density of a $r=R(\chi)$ hypersurface.
 Similarly we consider the pullback of $F$ to a hypersurface of constant $\chi$,
\begin{equation}
F{\bigr|}_{\chi=\mathrm{const}}=\rho_\chi\, dr\wedge d\phi\,,\qquad
\hbox{where}\quad \rho_\chi = \frac{\varepsilon \,q_m
\,\mathcal{A} \ell^2 G(\chi)}{r^2} \label{MaxPullBack2}
\end{equation}
is the charge density of a constant $\chi$ hypersurface.

We can now compute the charge of the several horizons. In view of
\eqref{lukeDoubleProperties}, hereafter we set $q_m=1/4$. Start with
the charge measured asymptotically outside the boundary black hole horizon, which is given by
\begin{equation}
\mathcal{Q}_{\sf  bdy}= \frac{1}{4\pi}\int_0^{\sqrt{2}\,\pi}d\phi
\int_{x_0}^{-b}d\chi \, \rho_R {\biggr |}_{R\rightarrow \infty} =
\frac{\ell}{32\sqrt{2}}\,(b-2)^2\sqrt{4+4b-b^2}\,. \label{Qbdy}
\end{equation}
This is the charge of the holographic fluid outside the horizon.
We can check that it matches the charge of the black funnel horizon.
For the black funnel of quadrant $II$, the charge includes a
contribution from the surface $FF'$ plus a contribution from the
surface $F'E$ in the left panel of Fig. \ref{fig:CoordDropFun}, and
yields
\begin{equation}
\mathcal{Q}_{BF}= \frac{1}{4\pi}\int_0^{\sqrt{2}\,\pi}d\phi
\left[\int_{x_0}^{-b}d\chi \, \rho_R {\biggr
|}_{R=-\frac{\mathcal{A}^2\ell^2 F(\chi)}{\chi+b}}
+\int_{\frac{4b(4-b)}{(2-b)(4+4b-b^2)}}^{\infty} dr\,\rho_\chi
 {\biggr |}_{\chi=-b}\right] = \mathcal{Q}_{\sf  bdy} \,.
\label{Qfunnel}
\end{equation}
(Notice that this horizon charge does not agree, as expected, with
the charge parameter $q_m$ of the black funnel.)

Consider now the black droplet system of quadrant $IV$. It has two
horizons (and associated charges): one is the horizon of the droplet
and the other of the deformed hyperbolic black hole. The charge of
the droplet (see $ED$ in the right panel of Fig.
\ref{fig:CoordDropFun}) and of the hyperbolic  (see $BC$ in the same
figure) horizons are, respectively,
\begin{eqnarray}
 &&\mathcal{Q}{\bigr |}_{y=-b}=
\frac{1}{4\pi}\int_0^{\sqrt{2}\,\pi}d\phi
\int_{\frac{1}{2(\sqrt{2}-1)+b}}^{\infty}dr \, \rho_\chi {\biggr
|}_{\chi=-b} = -\frac{\ell
 \left(4-2 \sqrt{2}+\sqrt{2} b\right) \left(8+4 b-6
b^2+b^3\right)}{64 \sqrt{4+4 b-b^2}}<0\,,\nonumber\\
 &&
\mathcal{Q}{\bigr |}_{y=b-4}=
\frac{1}{4\pi}\int_0^{\sqrt{2}\,\pi}d\phi \int_{x_0}^{\chi_C}d\chi
\, \rho_R {\biggr |}_{R=-\frac{\mathcal{A}^2\ell^2
F(\chi)}{b-4-\chi}}= \frac{\ell}{16}(2-b)\sqrt{4+4b-b^2}>0 \,.
 \label{QdropletAUX}
\end{eqnarray}
So the droplet and the deformed hyperbolic horizons have charges of
opposite sign such that their sum gives, as it should, the boundary charge (outside
the boundary black hole horizon),
\begin{equation}
\mathcal{Q}_{BD}= \mathcal{Q}{\bigr |}_{y=-b}+ \mathcal{Q}{\bigr
|}_{y=b-4} = \mathcal{Q}_{\sf  bdy} \,. \label{Qdroplet}
\end{equation}

This is an important observation that sheds light on the
interpretation of our results. Indeed, since the droplet and
deformed hyperbolic horizons have charges of opposite sign, these
two black holes have an extra electromagnetic attraction with
respect to the case where the horizons are uncharged. Due to this
extra attraction, we expect that the free energy is minimised by the
existence of a single horizon, and so the charged black funnels
should extend their regime of domination at the expense of the black
droplets. For the solutions that we studied, we find that the black
funnels actually dominate over the entire parameter range.

\setcounter{equation}{0}
\section{Holographic interpretation of black droplets and funnels \label{sec:AdSCFT}}

The stress tensor of the field theory that lives in the AdS boundary
of the black droplets and funnels is given by the
holographic stress tensor. This tensor is the starting point to
discuss the field theory description of the black droplets and
funnels in thermal equilibrium.

The induced stress tensor associated with the action
\eqref{totalAction} is given by the variation of the action with respect to the induced boundary metric
$\gamma^{ab}$, $\mathcal{T}_{ab}=\frac{2}{\sqrt{-\gamma}}\frac{\delta I}{\delta
\gamma^{ab}}$, yielding\footnote{This expression \eqref{sten1},
and \eqref{bdyStressT}, for the holographic tensor is derived from
the Euclidean action \eqref{totalAction} with the Fefferman-Graham
counterterms. This same expression also yields the
correct final answer for the stress tensor if we work with the
coordinates $\chi=x$, $y=x\pm1/r$, instead of \eqref{FGcoords}, even
though the action is then not finite. This is because the formula for the 
stress tensor is boundary covariant. A similar situation occurs
with the Kerr-Newman black hole: if \eqref{sten1} is evaluated in
the Boyer-Lindquist coordinate system, the correct stress
tensor is obtained notwithstanding that the action with the counterterm contribution \eqref{ActionContrib} diverges in this coordinate system
\cite{Papadimitriou:2005ii}.}
\begin{equation}
-8\pi\GN\, \mathcal{T}_{ab} = K_{ab} - \gamma_{ab}\,
K + \frac{2}{\ell} \, \gamma_{ab} - \ell\,
\left(\mathcal{R}_{ab} - \frac{1}{2}\, \mathcal{R}\,
\gamma_{ab}\right). \label{sten1}
\end{equation}
Since the induced boundary metric $\gamma_{ab}$ diverges near
the boundary, this induced stress tensor vanishes at the boundary.
In the standard AdS/CFT dictionary, the finite expectation value for
the holographic stress tensor is obtained by rescaling the induced
stress tensor by a power of the metric conformal factor,
\begin{equation}
\left<T_{ab}\right> = \lim_{r \to \infty}
\,-\frac{\varepsilon\,r}{\mathcal{A}}\, \mathcal{T}_{ab},
\label{bdyStressT}
\end{equation}
where $\varepsilon=+1$ in sectors II and III, and $\varepsilon=-1$
in sectors III and IV.\footnote{The $-\varepsilon$ factor guarantees
that the black funnel has the same $\left<T_{ab}\right>$ as the
black droplet and that this reduces to the $\left<T_{ab}\right>$ of
a perfect fluid asymptotically. This overall minus sign in the
conformal factor is irrelevant for the boundary metric
(\ref{boundaryMetric}) since one uses the square of the conformal
factor to obtain it.} After using the property $F'''(\chi) = 12\,
\mu+24(q_e^2+q_m^2)\chi$ of the polynomial $F$, we obtain
\begin{equation}
\left<T^a{}_b\right> = \frac{\mathcal{A}\ell^3}{8\pi\mathcal{G}}
\left[\mu +2(q_e^2+q_m^2) \chi \right] \,  \text{diag}\bigg\{
G(\chi) - 2 \, F(\chi),\, F(\chi) + G(\chi),\, F(\chi) - 2\,
G(\chi)\bigg\}. \label{Tupdn}
\end{equation}

In the Euclidean metric there is a symmetry between $x$ and $y$ and
this is reflected in the fact that the signs and factors of two flip
between the $tt$ and $\phi\phi$ components of $\left<T^a{}_b\right>$ as we pass
from the horizon to the axes. Note that due to the relation $F(\chi)
+ G(\chi) = \lp \mathcal{A}\ell\rp^{-2}$, the holographic stress
tensor is traceless, $\left<T^a{}_a\right>=h^{ab}\left<T_{ab}\right>=0$. This is a
consequence of the fact that the boundary field theory is odd
dimensional, and thus there is no conformal anomaly.

The form of the stress tensor \eqref{Tupdn} is that of a thermal gas
of massless radiation.  In general \eqref{Tupdn} does not correspond
to a perfect fluid,
$\left<T_{ab}\right>=P(\chi)(h_{ab}+3\,u_a u_b)$. However,
near the asymptotic region of the holographic boundary, $\chi\sim
x_0$, $G(\chi)$ vanishes and \eqref{Tupdn} does reduce to the stress
tensor of a perfect fluid. In this region, the boundary metric
approaches $\mathbb{R}\times\mathbb{H}^2$, and the bulk solution is
described by the geometry of the hyperbolic black hole
\eqref{ZoomCmetric}. The dual field theory for such a black hole has
been discussed in detail in \cite{Emparan:1999gf}. There, it was
shown that this black hole minimizes the free energy for all
temperatures and, from the field theory perspective, this solution
describes a deconfined thermal plasma that permeates the boundary.
Notice that this was probably an unexpected result since the
curvature of the hyperbolic space introduces a scale in the problem
that could potentially result in a critical temperature for a phase
transition between different solutions (this is what occurs for a
field theory on $\mathbb{R}\times S^2$, where the radius of $S^2$
sets a scale that introduces a critical temperature where a
confinement/deconfinement phase transition occurs \cite{Witten:1998qj}). The analysis of
\cite{Emparan:1999gf} was done only for  uncharged hyperbolic black
holes, but the main conclusions relevant for our purposes extend to
the charged case.

In the case at hand the black droplets and funnels only approach the
hyperbolic black hole in the asymptotic region $\chi\sim x_0$.
Moreover, they also introduce a boundary black hole. This means that
on the holographic boundary we have a boundary black hole in
equilibrium with a thermal fluid or plasma that fills all the
spacetime outside the black hole.

The difference between black funnels and black droplets is that, in
the latter case,  there are two disconnected event horizons in the
bulk, one associated to the boundary black hole, and the other one
associated to the asymptotic region of the boundary. The authors of
\cite{Hubeny:2009kz} propose the interpretation that  black funnels
describe boundary black holes that couple strongly to the deconfined
plasma. Under a thermal perturbation, heat flows easily from the
boundary black hole towards the plasma and thermal equilibrium is
rapidly achieved. In the bulk, such a perturbation flows
continuously from the boundary black hole down the throat of the
funnel and out of the shoulders of the funnel (described by the
hyperbolic black hole dual to the deconfined plasma). On the other
hand, the black droplet system describes a boundary black hole that
couples only weakly to the deconfined plasma. Under a thermal
perturbation it is now rather difficult for heat to flow from the
boundary black hole to the plasma. In the bulk description this is
because there is no connection between the droplet horizon (that
reaches the boundary horizon) and the hyperbolic horizon deep in the
bulk. As a consequence, the amount of heat that flows outward from
the boundary black hole resembles the flow that occurs in a confined
phase, even though the temperature is clearly much above any
potential deconfinement/confinement transition \cite{Hubeny:2009kz}.
From the field theory perspective, \cite{Hubeny:2009kz} interprets
the weak coupling between the black hole and plasma as being due to
the finite (and large) physical size of plasma excitations relative
to the black hole size.

A closer inspection into the expectation value $\left<T_{ab}\right>$ of the stress tensor given in (\ref{Tupdn}) shows that the energy
density $\left<T_{tt}\right>$ of the quantum fields living on the
boundary black hole background is positive in the asymptotic region,
but becomes negative close to the event horizon. This mirrors the behavior one obtains from free field calculations of Hawking radiation, and it is precisely this violation of the classical energy condition that allows the black hole to bypass the area theorem and evaporate through emission of thermal radiation once backreaction is taken into account (see for example the reviews \cite{Brout:1995rd}). We stress however that in our case no such backreaction occurs, because we have no propagating gravitational degree of freedom in the boundary theory. The results we obtain are not a `semi-classical' approximation as in usual Hawking radiation computations, but the exact quantum result.

The quantum fields propagating on the boundary black hole spacetime
behave  effectively as a fluid, with stress energy tensor given
precisely by (\ref{Tupdn}). Due to the bulk gauge field, in addition
to the background gravitational field described by the metric
(\ref{boundaryMetric}), there is also a background (non-dynamical)
electromagnetic field $F^{\sf bdy}_{ab}$, and the quantum fields
generate a current $J^a$. By the AdS/CFT dictionary,
 \eq F^{\sf
bdy}_{\mu\nu}=\lim_{r\rightarrow\infty}\gamma_\mu{}^\rho
\gamma_\nu{}^\si F_{\rho\si}\,,\qquad
\left<J^\mu\right>=\lim_{r\rightarrow\infty}\lp-\frac{\epsilon\,r}{{\mathcal
A}}\rp^3\gamma^{\mu\nu}n^\rho F_{\nu\rho}, \eeq
 yielding
  \eq F^{\sf
bdy}=q_e\ell\,\dd t\wedge\dd\chi+\frac{q_m}{\mathcal
A}\dd\chi\wedge\dd\phi\,,\qquad \left<J^a\right>=\left(-\frac{
q_e}{\mathcal A},0,q_m\ell\right). \eeq
It can be verified that these quantities verify the magnetohydrodynamic equation
\eq
D_a\left<T^a{}_b\right>=\frac1{4\pi\mathcal G}F^{\sf bdy}_{bc}\left<J^c\right>,
\eeq
and the current conservation equation
\eq
D_a\left< J^a\right>=0,
\eeq
as one expects on general grounds in AdS/CFT (here, $D_a$ is the boundary covariant derivative) \cite{mhd}. It follows that the Lorentz force $F^{\sf bdy}_{bc}\left<J^c\right>$ on the fluid elements is always centrifugal.
This force balances, together with the pressure of the fluid, the gravitational pull of the background geometry, leading to a stationary configuration. One peculiarity of the background geometry is that its constant time slices have finite volume, and so is the total charge carried by the fluid. Indeed, for the magnetic black droplet/funnel system ($q_e=0$), the total magnetic charge \eqref{Qbdy} measured asymptotically can be rewritten, by construction, as
\eq
{\mathcal Q}_{\sf bdy}=\frac1{4\pi}\int_{\mathcal E} F^{\sf bdy},
\label{3dQ}\eeq
where the integral is performed over the region $\mathcal E$ external to the black hole on a constant $t$ surface.
In three dimensions the flux \eqref{3dQ} is by definition the total magnetic charge carried by the plasma. This can be seen for example by performing an electromagnetic duality in the bulk, that transforms the magnetic solution in an electric solution, with electric charge $q_m$. On the boundary, the transformation induces a duality in which the roles of $F^{\sf bdy}$ and $\left< J^\mu\right>$ are swapped, through a three-dimensional Hodge duality. The dualized fields are now
\eq 
\tilde F^{\sf bdy}=\ast_3\left< J^a\right>=q_m\ell\,\dd t\wedge\dd\chi+\frac{q_e}{\mathcal
A}\,\dd\chi\wedge\dd\phi\,,\qquad 
\left<\vphantom{J^a}\right.\tilde J^a\left.\right>=\ast_3F^{\sf bdy}=
\left(\frac{q_m}{\mathcal A},0,q_e\ell\right), \eeq
and it follows that the flux integral \eqref{3dQ} can be recast as the volume integral of the magnetic charge density of the fluid,
\eq
{\mathcal Q}_{\sf bdy}=\frac1{4\pi}\int_{\mathcal E}\ast_3\left<J\right>=\frac1{4\pi}\int_{\mathcal E}\left<\right.\tilde J^a\left.\right>N_a\sqrt{\si}d^2\si, \label{BbdryHolo}
\eeq
where $N^a$ is the unit timelike normal to $\mathcal E$ in the boundary, and $\sqrt{\si}d^2\si$ its volume element. Therefore $\mathcal Q_{\sf bdy}$ is precisely the total magnetic charge of the boundary plasma, \ie the value of \eqref{BbdryHolo} agrees with \eqref{Qbdy}.

In conclusion, the phase transition predicted by Hubeny {\em et al.}
\cite{Hubeny:2009ru,Hubeny:2009kz,Hubeny:2009rc} for neutral boundary black holes does not occur when Maxwell fields are turned on, at least in the simple magnetically charged AdS C-metric system that we
have analyzed in this article. The reason is, that in a black droplet system the magnetic charges redistribute among the black droplet and the planar black brane with charges of opposite sign on the two horizons. The resulting extra electromagnetic interaction, that favors the merging of the horizons, raises the free energy of the black droplet system up to the point that the black funnel always dominates the partition function.

Hence, in this simple toy model for strongly coupled Hawking radiation, and for the known solutions of the theory that are in thermal equilibrium, the black funnel always dominates the canonical ensemble. 
We stress however that other saddle points, possibly with lower free energy, might exist and that, more importantly, our argument does not exclude such a phase transition for strongly coupled fields propagating on other boundary black holes, a four-dimensional Schwarzschild background, for instance. This will be checked only once the gravitational dual of such a field theory is built, presumably resorting to numerical methods.



\section*{Acknowledgments}

This article benefitted from several insightful discussions, that we
warmly acknowledge, with Roberto Emparan, Veronica Hubeny, Don
Marolf, Ji\v{r}\'{\i} Podolsk\'y, Mukund Rangamani, Harvey Reall
and Simon Ross. MMC was supported in part by the FWO-Vlaanderen,
project G.0235.05 and in part by the Federal Office for Scientific,
Technical and Cultural Affairs through the ``Interuniversity
Attraction Poles Programme -- Belgian Science Policy'' P6/11-P. MMC
acknowledges also support by the ANR grant STR-COSMO,
ANR-09-BLAN-0157, the ERC Advanced Grant 226371, the ITN programme
PITN-GA-2009-237920, the IFCPAR CEFIPRA programme 4104-2, the ANR
programme NT09-573739 ``string cosmo'' and the PEPS-CNRS programme
``Cordes, evolution, anisotropies et transitions''. OJCD
acknowledges financial support provided by the European Community
through the Intra-European Marie Curie contract PIEF-GA-2008-220197.
RM is funded by a postdoctoral fellowship of FNU-Denmark. This work
was partially funded by FCT-Portugal through project
PTDC/FIS/099293/2008.

\appendix

\section*{Appendices}


%
\renewcommand{\labelenumi}{\Roman{enumi}.}
\renewcommand{\labelenumii}{\Roman{enumi}.\Alph{enumii}.}

\section{The charged AdS C-metric: black droplets and black funnels  \label{sec:NoEquilibrium}}
\setcounter{equation}{0}

In this appendix, we shall explore the full parameter space of the charged AdS C-metric (\ref{AdSCmetric})-(\ref{AdSCmetricMaxwell}), and interpret the solutions in terms of black branes, black droplets and black funnels.
To simplify the discussion, we shall define the total charge parameter $q$ to be
\eq
q=\sqrt{q_e^2+q_m^2}\,,
\eeq
and use, following \cite{Hubeny:2009ru,Hubeny:2009kz}, the parameter $\la$ defined by
\eq
\la=\frac1{\mc{A}^2\ell^2}-1\,,
\eeq
in place of the acceleration parameter $\mc{A}$. Then, the solutions with negative cosmological constant we are interested in correspond to  the $\la>-1$ range.

As explained in Section~\ref{sec:Cmetric}, the asymptotic region of the manifold is located on the $x=y$ hypersurface, and the physical properties of the solutions are determined by the polynomials $F(\xi)$ and $G(\xi)$. In particular, roots of $F$ correspond to event horizons, and the zeroes of $G$ signal
the presence of axes of symmetry of the spacetime. Asymptotic regions on the boundary $x=y$ are located at double (or triple) roots of $G$. If such degenerate roots are absent, the boundary is compact.
In coincidence of these axes, we have a conical singularity unless the periodicity of the angular coordinate $\phi$ is set according to (\ref{PeriodPhi}).
If two such single roots are visible from a single static region (that is in the cases of a single black droplet, of a single planar black hole or of a black droplet floating on a planar black hole), there is a potential conical singularity, meaning that at least one of the horizons is accelerated by a cosmic string
stretching from it to infinity or to the other horizon, unless both singularities can be contemporaneously eliminated by a single choice of period for $\phi$, i.e. $|G'(x_i|)=|G'(x_j)|$.
Finally, the temperature of the event horizon at the root $y_i$ of the function $F(y)$ is obtained by requiring that no conical singularity appears in the Euclidean continuation of the manifold, and is given by (as explained in Section~\ref{sec:Cmetric})
\eq
T_i=\frac1\beta_i=\frac{|F'(y_i)|}{4\pi\sqrt{1+\la}}\,.
\label{T}\eeq
Again, when in presence of more than one horizon, a single choice of periodicity for the Euclidean time will not generally be sufficient to eliminate all conical singularities, and thermal equilibrium will be realized only for special cases.

Here, we want to classify the C-metric spacetimes according to their physical content in terms of black objects, and to this end we need to study parametrically the behavior of these polynomials, and then use the information on their roots to interpret the resulting spacetimes.

More precisely we are interested, given some static boundary conditions, in the static region of the bulk solution. It can present one or two event horizon; we will classify them according to the behavior of the $\p_\phi$ axes/singularities, following the definition proposed in \cite{Hubeny:2009kz}.
Since outer horizons lie at constant $y$, they intersect the $\p_\phi$ axes or singularities (that are at constant $x$) once or twice. If the horizon has two such intersections, it will obviously not reach the asymptotic boundary: such black holes are referred to as {\em planar black holes}, irrespectively of their
geometry or topology. If the horizon intersects a single axis/singularity, two situations can arise; if the axis connects the horizon to the boundary through the external region, the horizon is called a {\em black funnel}, while if the connection occurs in its interior, it is dubbed {\em black droplet}.

To identify the black hole content of a particular solution,
determined by a choice of physical parameters, it is convenient to
represent on the $(x,z)$ plane (with $z=y-x$) the roots $y=y_i$
corresponding to the horizons, and the roots $x=x_i$ corresponding
to the axes (or asymptotic regions of the boundary if they are
degenerate; in that case we represent them with double lines). An
example, corresponding to the thermal equilibrium configuration with
an asymptotic region studied in this article, is given in
Fig.~\ref{fig:DiagDropFun}. One then determines the static and
non-static regions, that are separated by the $y_i$ roots; in our
diagram we shaded the non-static regions. Then, each connected white
region corresponds to a single static solution of the bulk
gravitational equations. The dashed horizontal line at $x=y$
represents the boundary of the asymptotically AdS spacetime. If two
different static regions share the same boundary at $x=y$, then they
represent two competing gravitational saddle points dual to the same
field theory state. Then, the interpretation of the horizons present
in the solution according to the definition given in
\cite{Hubeny:2009kz}, is easily obtained according to Fig.~\ref{fig
interpr}.
\begin{figure}[h]
\centerline{\includegraphics[width=.95\linewidth]{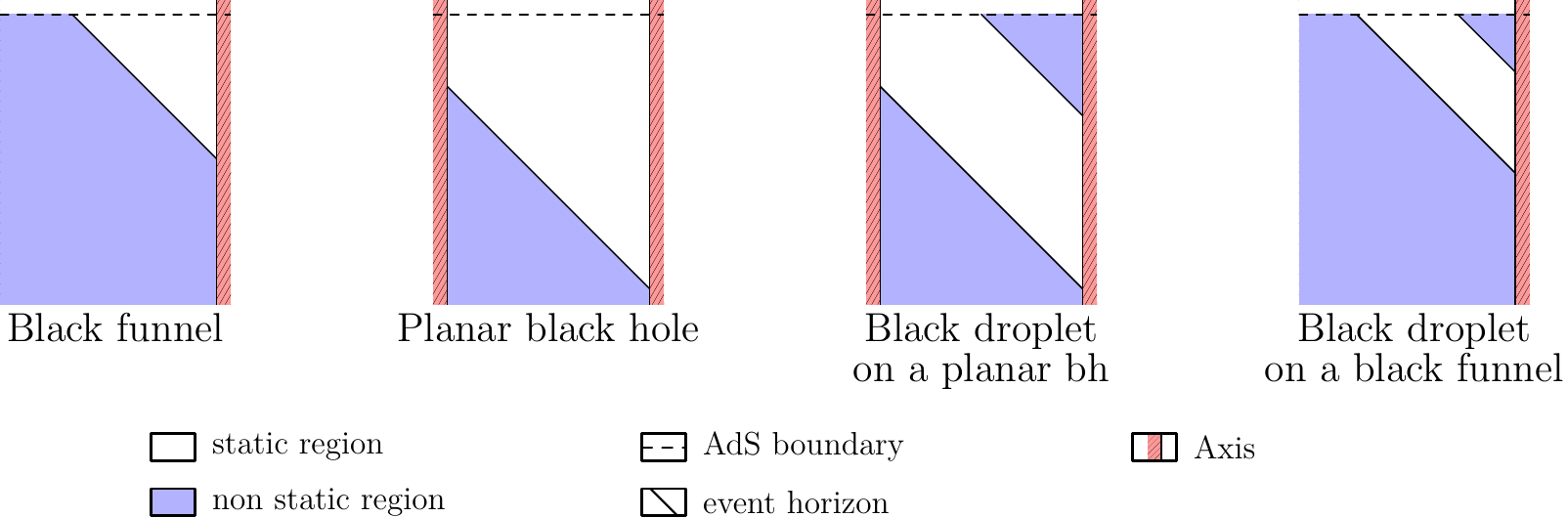}}
\caption{\small {\sc Interpretation of the solutions:} possible kinds of static regions that can be seen from a static boundary region, interpreted in terms of black droplets, black funnels, and planar black holes. The planar black hole and the black droplet floating over a planar black hole cases have two zeros of
the function $G$ in the static region, and therefore suffer in general of a conical singularity. Other cases can arise where one of the $x$ constant boundaries is a double or triple root of $G$, in which case it corresponds to an asymptotic region of the boundary, and black funnels and black droplets with
degenerate horizons.}
\label{fig interpr}
\end{figure}
The interested reader can find more details on the procedure in \cite{Hubeny:2009kz}.

We will consider the three cases $\kappa=0,\pm1$ separately.

\subsection*{$\kappa=-1$, hyperbolic horizons}

The function $G(\xi)$ takes in this case the form $G(\xi)=1+\xi^2-2\mu\xi^3-q^2\xi^4$, whose derivative has exactly three single real roots, in $\xi=0$ and in
\eq
\tilde x_{0,1}=-\frac{1}{\sqrt2}\frac{q_\star}{q^2}
\lp1\pm\sqrt{1+\frac{q^2}{q^2_\star}}\rp\,,
\eeq
then $\tilde x_0<0<\tilde x_1$, and $G$ has local maxima in $\tilde x_{0,1}$ and a minimum in the origin. It is furthermore easy to show that $G(\tilde x_0)\geq G(\tilde x_1)>1$ with the equality holding only for $\mu=0$. Therefore $G$ has precisely two single real roots $x_0$ and $x_1$, such that $x_0<
\tilde x_0<0<\tilde x_1<x_1$.
As we shall see, the root structure is the same as that arising in the $\kappa=+1$ III.A. case.
Varying $\la$, the roots merge and split, leading to static regions with black funnels/droplet/branes and singularities, that can easily be recognized by drawing the corresponding $x$-$z$ diagrams.

\subsection*{$\kappa=0$, flat planar horizons}

The function $G(y)$ has a maximum in $\tilde y_0=-\frac{3\mu}{2q^2}<0$ and an inflection point in $\tilde y_1=0$ corresponding to a double root of its derivative. $G$ has therefore two real roots $y_0$ and $y_1$, satisfying the relation $y_0<\tilde y_0<0<y_1$.
Hence, the function $F(\xi)$, obtained by a reflection of $G(\xi)$ with respect to the $\xi$ axis, followed by a translation, has a minimum in $\tilde y_0$, and zeroes if $F(\tilde y_0)\leq0$. Hence, defining
\eq
\la_\star=\frac{27\mu^4}{16q^6},
\eeq
$F$ has two real roots $x_0$ and $x_1$ for $-1<\la<\la_\star$, one double root $x_0=x_1$ for $\la=\la_\star$ and no roots if $\la>\la_\star$.
The root structure is the same as for cases $\kappa=+1$ I, II.A, III.D (with $x_2$ assuming the role of $x_0$) and III.E, that we shall study shortly. The static region has a conical singularity for $\la\geq\la_\star$, otherwise it represents a regular black funnel or a black droplet with a naked singularity.

\subsection*{$\kappa=+1$, spherical horizons}

The derivative of the function $G(\xi)$ is now
\eq
G'(\xi)=-4q^2\xi P(\xi)\,,\qquad
P(\xi)=\xi^2+\frac{3\mu^2}{2q^2}\xi+\frac{1}{2q^2},
\eeq
and the behavior of the second-order polynomial $P(x)$ is determined by the relative value of $q$ with respect to the critical charge
\eq
q_\star=\frac{3\mu}{2\sqrt{2}}.
\eeq
In particular $P(\xi)$ has no real roots for $q>q_\star$, one double root $\tilde x=-2/3\mu$ for $q=q_\star$, and two negative roots $\tilde x_0<\tilde x_1<0$ for $q<q_\star$.
\begin{enumerate}
\item $q>q_\star$

In this case $P(\xi)>0$ and $G'(\xi)$ has a single root in $\xi=0$, where $G(\xi)$ assumes its maximum. Therefore, $G(\xi)\leq1$ and $F(y)$ has two single real roots $\tilde y_0<0<\tilde y_1$ for $-1<\la<0$, one double real root in $\tilde y_0=0$ for $\la=0$ and no real roots for $\la>0$.
Again, with the same root structure as in the $\kappa=0$ case, we have conical singularities for $\la\geq0$, and otherwise black funnels or a black droplet with a naked singularity.

\item $q=q_\star$

$P$ has a double root in $\tilde x=-2/3\mu$, where $G(x)$ has an inflection point. Therefore, the function $G$ has precisely two roots $x_0$ and $x_1$. If we define the critical mass $\mu_\star=\sqrt{2}/3\sqrt{3}$, then we have the following cases:
\begin{enumerate}
\item $q_\star\neq\frac1{2\sqrt3}$ ($\mu\neq\mu_\star$)

then both real roots $x_1$ and $x_2$ are real roots, and they satisfy
$x_0<\tilde x<0<x_1$ for $\mu>\mu_\star$ and
$\tilde x<x_0<0<x_1$ for $\mu<\mu_\star$. The behavior is hence the same as the previous case I and as the $\kappa=0$ one with conical singularities and singular black droplets, apart from a very special case discussed below ($\mu<\mu_\star$ and $\lambda=\lambda_\star$).
\item $q_\star=\frac1{2\sqrt3}$ ($\mu=\mu_\star$)

In the critical case $\mu=\mu_\star$, the smaller root of $G$, $x_0=\tilde x=-\sqrt6$ becomes a triple root, and the polynomial becomes
\eq
G(x)=-\frac1{12}\lp x-\sqrt{\frac23}\rp\lp x+\sqrt6\rp^3.
\eeq
Now the solution develops a new asymptotic region on the boundary in correspondence of the first axis that gets pushed infinitely far away from any other point in the manifold, while the second axis $x_3$ is smooth as we choose the associated periodicity for the angle $\varphi$. The boundary field theory
states are therefore dual to two competing gravitational saddle points, one representing a regular black funnel, the other a single black droplet. Unfortunately, the black droplet has a curvature singularity in $y\rightarrow\infty$ that is not clothed by any horizon.

\end{enumerate}
Since $G(x)\leq1$ for any $x$, $F$ has no roots for $\la>0$, and a double root $y_0=y_1=0$ for $\la=0$. Finally, if $-1<\la<0$ the polynomial $F$ has two real roots $y_0$, $y_1$ satisfying $x_0<y_0<0<y_1<x_1$. These roots are always single roots for $\mu\geq\mu_\star$,
but when $\mu<\mu_\star$ the smaller root $y_0$ becomes a triple root for the critical value $\lambda=\la_\star=-\mu^2_\star/\mu^2=-1/12q_\star^2$. The different cases are therefore:
\begin{itemize}
\item $\lambda>0$:
No roots of $F$ (no horizons) $x_0<0<x_1$
\item $\lambda=0$:
One double root of $F$ in $y=0$, $x_0<y_0=y_1=0<x_1$
\item $-1<\lambda<0$:
Two roots $y_{0,1}$ of $F$, $x_0<y_0<0<y_1<x_1$. $y_1$ is a single root, $y_0$ is a triple root for $\mu<\mu_\star$, $\la=\la_\star$, and a single root otherwise.
\end{itemize}
\item $q<q_\star$

In this case $P(\xi)$ has two single real roots $\tilde x_0<\tilde x_1<0$, in correspondence of which the function $G(x)$ assumes a maximum and a minimum respectively. If $q>1/2\sqrt{3}$, both $G(\tilde x_0)$ and $G(\tilde x_1)$ are positive, and $G(x)$ has only two single roots. If $q<1/2\sqrt{3}$, define
\eq
q^\pm_\star(q)=\frac1{4\sqrt{3}}\sqrt{1+36q^2\pm\left(1-12q^2\right)^\frac32},
\eeq
such that $G(\tilde x_0)=0$ for $q^-_\star=q_\star$ and $G(\tilde x_1)=0$ for $q^+_\star=q_\star$.
Then, we have the following cases
\begin{enumerate}
\item $q_\star>1/2\sqrt{3}$ or $q_\star>q^+_\star$

both $G(\tilde x_0)$ and $G(\tilde x_1)$ are positive, and $G$ has two single real roots. We are therefore qualitatively in the same situation as that described in the $\kappa=-1$ case.

\item $q_\star=q^+_\star$

$\tilde x_1$ is a double root of $G$, hence $G$ has two three real roots $x_0<x_1<0<x_2$, with $x_0$ and $x_2$ single roots and $x_1=\tilde x_1$ double root. Note that in correspondence of the double root of $G$ there is an asymptotic region, and its intersection with the $z=0$ boundary gives the
asymptotic region of the boundary. The solutions in thermal equilibrium on which we focus in this paper belong to this class, for special parameters. For illustrative purposes, we show all the diagrams in the $z$-$x$ plane in fig.~\ref{zx IIIB} for this class of solutions. On them, one can observe the evolution of
the horizons as the parameter $\lambda$ is varied. Similar behavior is obtained for the other cases, although we choose not to show all the diagrams to keep the presentation short.
\begin{figure}[tcbh]
\centerline{\includegraphics[width=.96\linewidth]{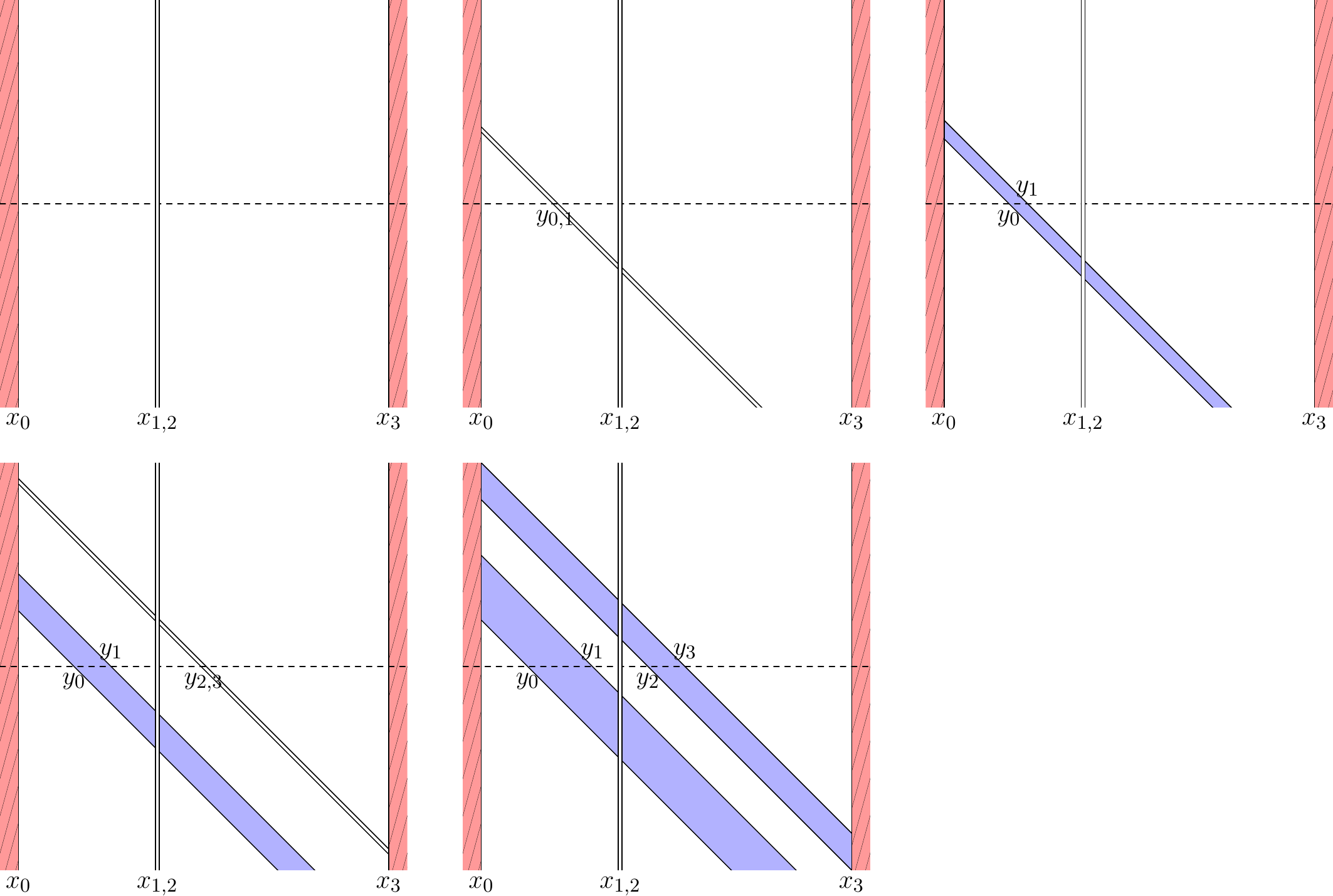}}
\caption{\small Diagram of the $z$-$x$ plane corresponding to the case III.B. The double root $x_1=x_2$ is the asymptotic region, and its intersection with the $z=0$ dashed line represents the asymptotic region of the boundary.
}
\label{zx IIIB}
\end{figure}
\item $q^-_\star<q_\star<q^+_\star$

the local minimum in $\tilde x_1$ is negative, while the maximum in $\tilde x_1$ is positive, hence $G$ has four single real roots $x_0<\tilde x_0<x_1<\tilde x_1<x_2<0<x_3$. Now the spacetime has axes of symmetries in both $x_1$ and $x_2$, and exhibits generally a conical singularity that cannot be
eliminated.

\item $q_\star=q^-_\star$

now $\tilde x_0$ is a double root of $G$, and we have three roots $x_0=\tilde x_0<x_1<0<x_2$, with $x_{1,2}$ single roots and $x_0$ double root. Since the double root appears in a region with wrong signature, it is not relevant for us and the allowed range of the $x$ coordinates is simply $(x_2,x_3)$. The
resulting $z$-$x$ diagrams are therefore the ones already described in the $\kappa=0$ case, where now $x_2$ has the role of $x_0$.

\item $q<q_\star<q^-_\star$

both local extrema of G in $\tilde x_{0,1}$ are negative, therefore $G$ has only two single real roots in $x_0<0<x_1$. The situation is again analogous to the one described in the $\kappa=0$ case.
\end{enumerate}

We will not enter in the details of what happens as $\la$ is varied, as this can be easily determined case by case, but the situation can be summed up by saying that the roots $y_2$ and $y_3$ of $F$ are always present for $-1<\la<0$, they merge in a double root for $\la=0$ and they disappear for $\la>0$.
The roots $y_0$ and $y_1$ exist as long as $\la<G(\tilde x_0)-1$. For the critical value $\la_\star=G(\tilde x_0)-1$, they merge in a double root, and for $\la>\la_\star$ they disappear.

In the case III.A., in addition of the previous behavior of the roots of $F$, one has that for $\la=G(\tilde x_1)-1$ the roots $y_1$ and $y_2$ merge, and they disappear for lower values of $\la$.
\end{enumerate}

\begin{figure}[tcbh]
\centerline{\includegraphics[width=.46\linewidth]{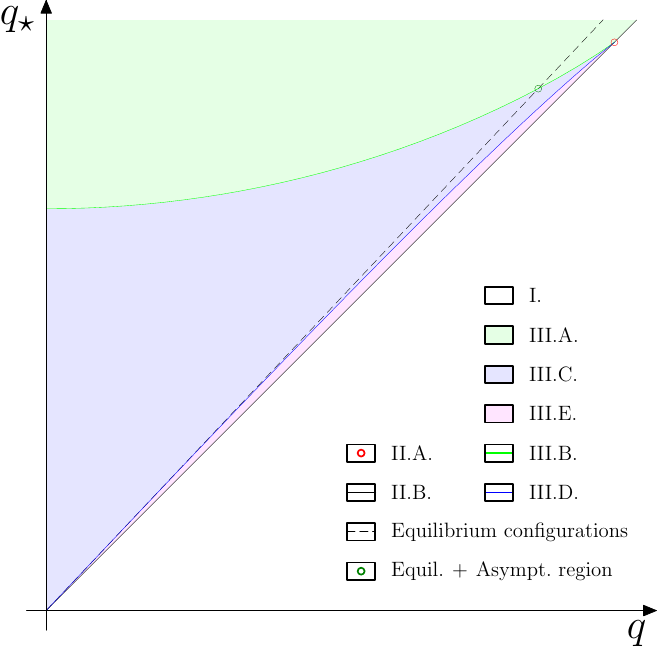}}
\caption{\small Summary for the $\kappa=+1$ case}
\label{figk1summary}
\end{figure}

The different behaviors of the polynomial $G$ have been summarized in fig.~\ref{figk1summary} in the parameter space $q_\star$-$q$ (or equivalently the $\mu$-$q$ plane).

\section{Configurations in thermal equilibrium\label{appB}}

As we discussed in the previous section, the C-metric presents an event horizon at each root $y_i$ of the function $F(y)$, whose temperature is given by equation (\ref{T}).
It follows that generically, solutions with more than one event horizons are not in thermal equilibrium because the corresponding black holes are at different temperatures. In particular, black droplets floating over planar branes (or over black funnels) have different temperatures and, taking into account
quantum corrections, we cannot expect them to be true stationary equilibrium configurations; the evolution should drive them through quantum evaporation to the funnel solution with the same temperature as the initial black droplet solution (fixed by the boundary conditions). It is therefore interesting to look
whether such equilibrium configurations can exist. To this end, we parameterize the polynomial $F(y)$ using its roots $y_0,\ldots,y_3$, and assume that at least $y_1<y_2$ are distinct real roots. Then
\eq
F(y)=q^2\prod_{i=0}^3(y-y_i)=q^2\lp y^4-\si y^3+t y^2-\rho h y+\rho\rp
\eeq
with
\eq
\si=\sum y_i=-\frac{2\mu}{q^2}\,,\quad
\rho=\prod y_i=\frac{\kappa}{q^2}\,, \quad
h=\sum\frac1{y_i}=0\,,\quad
t=\sum_{i<j}y_i y_j=\frac{\la}{q^2}\,.
\eeq
The equilibrium requirement that the horizons in $y_0$ and $y_1$ have the same temperature is then $|F'(y_1)|=|F'(y_2)|$; however if $F'$ has the same sign on both roots, there must be a third root in between, and $y_1$, $y_2$ are not separated by a static region. We will therefore look for values of the
parameters such that $F'(y_1)=-F'(y_2)$. This condition implies that the other two roots are real and satisfy $y_0+y_3=y_1+y_2$ and $y_0y_3=-y_1y_2$, which can be solved to give
\eq
y_{0,3}=\frac12\lp y_1+y_2\mp\sqrt{(y_1+y_2)^2+4y_1y_2}\rp.
\eeq
Therefore, configurations in thermodynamic equilibrium always have four horizons, ordered according to $y_0<y_1<y_2<y_3$. Denoting $y_1=-a$ and $y_2=-b$, the polynomial can be rewritten in the form
\eq
F(y)=-\lp\frac{ab}{a+b}\rp^2\lp1+\frac y{a}\rp\lp1+\frac y{b}\rp\lp1-\frac{a+b}{ab}y-\frac{1}{ab}y^2\rp.
\eeq
Then, $\kappa=q^2(a+b)^2>0$ and therefore we are in the $\kappa=+1$ case, and the physical parameters can be written parametrically in terms of $a$, $b$ as
\eq
\mu=q=\frac1{a+b}\,,\qquad
\la=\frac1{\ell^2\mc A^2}-1=-\lp\frac{ab}{a+b}\rp^2\,,
\eeq
with the restriction $a+b>0$ coming from positivity of $\mu$ and $q$.
The critical charge reads then $q_\star=3q/{2\sqrt2}$.
We are therefore in the case $\kappa=+1$, III ($q<q_\star$) of our classification, with $q_\star>q_\star^-$ as can be readily checked; and the value of $\mu=q$ fully determines the black objects content of the solution, as follows:
\begin{itemize}
\item
$\mu=q>1/4$:\ Case III.A, we have a black funnel at thermal equilibrium with a black droplet floating over it. The asymptotic boundary is conformally compact and the static region of the metric is regular.
\item
$\mu=q=1/4$:\ Case III.B, the condition implies $a=4-b$, and results in the one parameter family of solutions given explicitely in equation  (\ref{lukeDoubleProperties}). The boundary is not compact and has an asymptotic region, the solution represents a planar black hole at equilibrium with a black droplet
floating over it in one sector, and a black funnel in the other. This is the regular solution that was analysed in detail in Sections \ref{sec:DropletFunnelEquilibrium} and \ref{sec:FreeEtensor}.
\item
$\mu=q<1/4$:\ Case III.C, again a planar black hole at equilibrium with a black droplet floating over it, but with a conformally compact boundary. Potentially a conical singularity could occur in this case, but a simple check shows that they cannot be eliminated simultaneously on both axis if $0<\mu<1/4$ and
therefore this case is singular.
\end{itemize}
Indeed, the roots of the function $G(x)$ that signals the axes of symmetry are given explicitely by
\eq
x_0=\frac{-1-\sqrt{1+4\mu}}{2\mu}\,,\quad
x_1=\frac{-1-\sqrt{1-4\mu}}{2\mu}\,,\quad
x_2=\frac{-1+\sqrt{1-4\mu}}{2\mu}\,,\quad
x_3=\frac{-1+\sqrt{1+4\mu}}{2\mu}\,.
\eeq
The roots $x_1$ and $x_2$ do not exist for $\mu>1/4$, for $\mu=1/4$ they degenerate in a double real root corresponding to the asymptotic region of the boundary, and for $0<\mu<1/4$ we have four distinct single real roots $x_0<x_1<x_2<x_3$. It is then trivial to check that the relation $G'(x_2)=-G'(x_3)$
(or $G'(x_0)=-G'(x_1)$) cannot be satisfied for any choice of $0<\mu<1/4$, meaning that a conical singularity is unavoidable.

\begin{figure}[tcbh]
\centerline{\includegraphics[width=.96\linewidth]{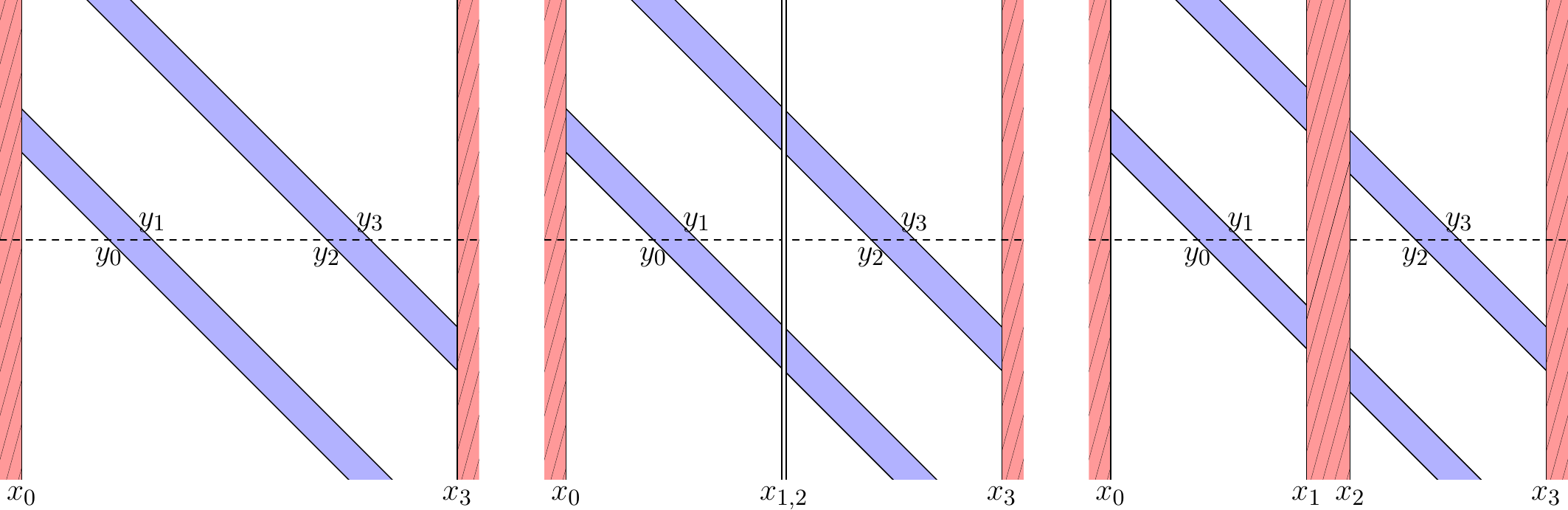}}
\caption{\small Equilibrium configurations with $\kappa=+1$, $\mu=q$ and $\lambda<0$. The horizons share the same temperature, and the spacetime is therefore at thermal equilibrium. From left to right, a black droplet floating over a black funnel for $\mu>1/4$, a black droplet floating on a black hole with
an asymptotic region for $\mu=1/4$, and a black droplet and a planar black hole with a conical singularity for $\mu<1/4$.}
\label{equil2}
\end{figure}

If $\lambda=0$, the two horizons are degenerate, corresponding to double roots of $F$, and the temperature of the system vanish. For $-1<\la<0$, the horizons have finite temperature and the solutions correspond to systems of two horizons at thermal equilibrium. The corresponding $z$-$x$ diagrams can
be found in fig.~\ref{equil2}.

These equilibrium configurations are plotted with the dashed lines in fig.~\ref{figk1summary}.
The first two cases have genuine regular external regions at thermal equilibrium, the latter having been investigated thoroughly in the body of this article. The reader might wonder whether a similar analysis, involving two geometries competing to minimize the free energy could be performed in the first
case. Before trying to answer this question, we need however to establish a symmetry of those solutions.
\paragraph{A symmetry relating configurations in thermal equilibrium:} the requirement that the temperatures of the horizons at $y_0$ and at $y_1$ coincide imposes that the derivatives of the polynomials $F$ and $G$ in those two points are opposite and the tangents to these curves are reflected through a
vertical axis. This in turn forces the full polynomials to be symmetric under that same reflection, and induces a discrete isometry on the C-metric.
Indeed, if one shifts the coordinates $x$ and $y$ according to
\eq
\tilde x=x+\frac1{2\mu}\,,\qquad\tilde y=y+\frac1{2\mu}\,,
\eeq
the metric (\ref{AdSCmetric}) remains invariant in form since $\tilde x-\tilde y=x-y$. The $F$ and $G$ polynomials get transformed into
\eqn
&&\tilde F(\tilde y)=F(\tilde y-\frac1{2\mu})=\frac1{(a+b)^2}\lb \tilde y^2-\lp\frac{a-b}2\rp^2\rb
\lb \tilde y^2-\frac14\lp a+b\rp^2-ab\rb,\\
&&\tilde G(\tilde x)=G\lp\tilde x-\frac1{2\mu}\rp=-\frac{\tilde
x^4}{(a+b)^2}+\frac12\tilde x^2+\frac2{\A^2\ell^2}
-\frac{(a-b)^2}{16(a+b)^2}. \eeqn These polynomials are even, and
therefore the full metric (\ref{AdSCmetric}) is invariant left
invariant under the transformation \eq \tilde x\mapsto-\tilde
x\,,\qquad \tilde y\mapsto-\tilde y\,. \label{parity}\eeq On the
other hand, the gauge potential transforms as $A\mapsto A$, and the
electric charge and magnetic charge get reversed under this
transformation, $q_e\mapsto-q_e$ and $q_m\mapsto-q_m$. This isometry
acts therefore on the $\mu=q=1/4$ solution by exchanging regions $I$
and $III$ with regions $IV$ and $II$ respectively in diagram of
Fig.~\ref{fig:DiagDropFun} respectively, and inverting the charges;
we conclude that region $I$ is isometric to region $IV$ and region
$II$ is isometric to region $III$, as advertised in
section~\ref{sec:LukeCmetric}, and (up to the sign of the charges)
the results obtained for quadrant $I$ ($III$) are valid also for
quadrant $IV$ ($II$).

\begin{figure}[tcbh]
\centerline{\includegraphics[width=.32\linewidth]{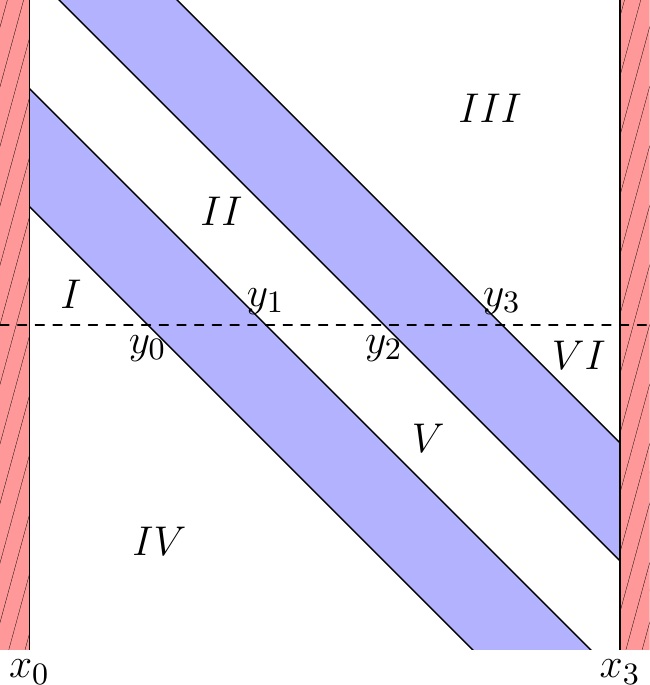}}
\caption{\small Thermal equilibrium configuration with $\kappa=+1$ and $\mu=q>1/4$, case III.A.
The regions $II$ and $V$ share the same boundary, but they are related by an isometry of the C-metric. }
\label{symm}
\end{figure}

We finally can consider the $\mu=q>1/4$. The corresponding diagram
is shown in Fig.~\ref{symm}, and has six distinct static external
regions. Of those, only regions $II$ and $V$ are regular, but as
they share the same asymptotic boundary, one might wonder which one
dominates the partition function of the dual CFT. Again, the
isometry (\ref{parity}) comes to our help since it maps region $II$
to region $V$, and therefore these regions do not represent two
distinct gravitational instantons in competition to minimize the
free energy, and no black funnel/black droplet transition is present
in this case.

In conclusion, the $\mu=q=1/4$ case studied in this article is the only one in which black droplets and black funnels at thermal equilibrium compete to dominate the free energy among the solutions of the charged AdS C-metrics.


\end{document}